\documentclass[preprint]{aastex} 
\usepackage{color}

\newcommand{\referee}[1]{#1}

\newcommand{\tab}[1]{Table~\ref{#1}}
\newcommand{\fig}[1]{Figure~\ref{#1}}
\newcommand{\eq}[1]{Equation~(\ref{#1})}
\newcommand{\sect}[1]{Section~\ref{#1}}

\newcommand{\lgamma}{L_{\gamma}}
\newcommand{\lgammaNorm}{L_{\gamma, 52}}
\newcommand{\egamma}{E_{\gamma}}
\newcommand{\egammaNorm}{E_{\gamma, 52}}

\newcommand{\tpobs}{T_{\rm p}^{\rm obs}}
\newcommand{\tpz}{T_{\rm p}}
\newcommand{\tburst}{T_{90}}
\newcommand{\tgamma}{T_{\gamma}}

\newcommand{\dtobs}{\Delta t_{\rm obs}}
\newcommand{\tdec}{T_{\rm dec}}

\newcommand{\rp}{R_{\rm p}}
\newcommand{\rph}{R_{*}}
\newcommand{\rdec}{R_{\rm dec}}
\newcommand{\rhodec}{\rho_{\rm dec}}
\newcommand{\rdiss}{R_{\rm diss}}

\newcommand{\ep}{E_{\rm p}}
\newcommand{\ecut}{E_{\rm max}}

\newcommand{\lfbw}{\Gamma_{\rm bw}}
\newcommand{\lfej}{\Gamma}

\newcounter{refcompteur}
  \def\generateur#1{%
  \begingroup
  \edef\next{\def\expandafter\noexpand\csname #1\endcsname{\therefcompteur}}%
  \expandafter\endgroup\next
  \addtocounter{refcompteur}{1}
}

\newcommand{\reftab}[2]{(#1) \citealt{#2}}

\shorttitle{Lorentz factors of gamma-ray bursts} 
\shortauthors{Hascoet et al.}

\begin{document}

\title{Estimates for Lorentz factors of gamma-ray bursts \\
from early optical afterglow observations}

\author{Romain Hasco\"et\altaffilmark{1}, Andrei M. Beloborodov}
\affil{Physics Department and Columbia Astrophysics Laboratory, Columbia University, 538 West 120th Street New York, NY 10027}

\author{Fr\'ed\'eric Daigne, Robert Mochkovitch}
\affil{Institut d'Astrophysique de Paris, UMR 7095 Universit´e Pierre et Marie Curie - CNRS, 98 bis boulevard Arago, Paris 75014, France}

\altaffiltext{1}{\texttt{hascoet@astro.columbia.edu}}

\begin{abstract}
The peak time of optical afterglow may be used as a proxy to constrain the Lorentz factor $\lfej$ of the gamma-ray burst (GRB) ejecta.
We revisit this method by including bursts with optical observations that started when the afterglow flux was already decaying;
these bursts can provide useful lower limits on $\lfej$.
Combining all analyzed bursts in our sample, we find that the previously reported correlation between $\lfej$ and the burst luminosity $\lgamma$ does not hold.
However, the data clearly shows a lower bound $\lfej_{\rm min}$ which increases with $\lgamma$.
We suggest an explanation for this feature: explosions with large jet luminosities and $\lfej < \lfej_{\rm min}$ suffer strong adiabatic cooling 
before their radiation is released at the photosphere; they produce weak bursts, barely detectable with present instruments.
To test this explanation we examine the effect of adiabatic cooling on the GRB location in the $\lgamma - \lfej$ plane 
using a Monte Carlo simulation of the GRB population.
Our results predict detectable on-axis ``orphan'' afterglows.
We also derive upper limits on the density of the ambient medium that decelerates the explosion ejecta.
We find that the density in many cases is smaller than expected for
stellar winds from normal Wolf-Rayet progenitors. 
The burst progenitors may be peculiar massive stars with weaker winds or there might 
exist a mechanism that reduces the stellar wind a few years before the explosion.
\end{abstract}


\section{Introduction}
\label{sect_intro}

The prompt emission of gamma-ray bursts (GRBs) is likely produced by dissipative mechanisms inside the relativistic ejecta of the explosion, 
while the GRB afterglow is associated with 
the ejecta deceleration by a circum-burst medium (see e.g. \citealt{piran_2004} for a review).
The afterglow emission is attributed to a relativistic blast wave that involves a pair of shocks -- forward and reverse.

One of the most important parameters of GRBs is the Lorentz factor of the relativistic ejecta, $\lfej$, as the models of the prompt and afterglow emissions strongly depend on its value.
Useful constraints on
$\lfej$ may be derived using the timescale and spectrum of the prompt GRB \citep{paczynski_1986, goodman_1986}, which typically gives $\lfej > 100$ (e.g. \citealt{lithwick_2001, hascoet_2012}).
Another possible way to constrain the GRB Lorentz factor is to use the afterglow peak time $\tpz$ as a proxy for the deceleration time of the blast wave, 
$\tdec$, 
at which the dissipation rate peaks (e.g. \citealt{meszaros_1997, sari_1999b}).
Using this method \citet{liang_2010, liang_2012} and \citet{ghirlanda_2012} studied a sample of GRBs with detected 
optical peaks and found 
a correlation between $\lfej$ and the burst luminosity $\lgamma$.

The key assumption of this method,
$\tpz \sim \tdec$, is questionable, as optical emission could reach its peak at a different time.
This possibility is illustrated by the simple model of synchrotron emission from a self-similar blast-wave, where the optical light curve can peak at $\tpz \gg \tdec$ (e.g. \citealt{sari_1998}).
However, observations conflict with the late-$\tpz$ models and lend some support to the 
$\tpz \sim \tdec$ assumption.
In many bursts, optical  emission peaks early and steeply, as may be expected at $\tdec$.\footnote{This expectation depends on the 
model for the circum-burst density. A steep rise is 
firmly predicted if the density is uniform, $\rho=const$, but 
questionable if $\rho\propto R^{-2}$ (wind-type medium, \citealt{chevalier_2000}).
In the latter case, additional effects such as $e^\pm$ pair loading 
could produce the steep rise toward the peak \citep{beloborodov_2002}.}
The optical peak may be dominated by the forward- or reverse-shock emission 
(e.g. \citealt{meszaros_1997, sari_1999b, uhm_2007, genet_2007}). 
In this paper, we accept $\tpz \simeq \tdec$ as a reasonable assumption and investigate its implications.

In Sections~2 and 3, we extend the previous 
analysis  
by including bursts whose afterglow peaked before observations started,
which provides a useful upper limit on $\tpz$. We also identify the
cases where the blast wave at $\tpz$ is significantly slower than the ejecta, which corresponds to a relativistic reverse shock;
in these cases the measurement of $\tpz$ provides only a lower bound on the ejecta Lorentz factor $\lfej$.
Our analysis does not support the existence of the $\lgamma - \lfej$ correlation claimed in previous studies.
Instead, the data shows a lack of bright bursts with low Lorentz factors.
In Section~4, we suggest an explanation of this fact.
Section~5 summarizes our results and discusses implications of observed $\tpz$
for the nature of the circum-burst medium.

\section{Sample}
\label{sample}

\subsection{Bursts with detected optical afterglow peaks}

\tab{tab_detec} gives the list of \referee{20} GRBs with \textit{detected} early afterglow peaks 
that are included in our sample.
This list is a selection from the GRB samples of \citet{liang_2010, lue_2012, liang_2012}, where we keep only bursts with reliable detection of the peak time. We removed 
(1) GRBs that have optical light-curves with more than one bump, making the peak measurement ambiguous, 
(2) GRBs for which the peak was measured during a plateau phase (i.e. where the optical light-curve is flat in logarithmic scale), and 
(3) GRBs with optical light-curves that are sampled too sparsely or whose temporal range is too small to provide significant constraints on $\tpz$. 

The redshift-corrected  peak times $\tpz = \tpobs/(1+z)$
 are shown in \fig{fig_obs} versus the GRB isotropic equivalent gamma-ray energy $\egamma$ and the corresponding average luminosity 
 \begin{equation}
 \label{eq_lum}
 \lgamma = \frac{\egamma}{\tgamma} \, , 
 \end{equation} 
 where 
 \begin{equation}
 \label{eq_lum}
 \tgamma = \frac{\tburst}{1+z} , 
 \end{equation} 
 and $\tburst$ is an approximate measure of the observed burst duration (time during which $90$\% of the emission is received). 
One can notice a good correlation between $\tpz$ and $\lgamma$ (or $\egamma$);
we argue below that this correlation is spurious.

\setcounter{refcompteur}{1}
\generateur{liangA}
\generateur{lueC}
\generateur{liangC}
\generateur{molinariG}
\generateur{stamatikosH}
\generateur{martinI}
\generateur{guidorziB}
\generateur{yuanA}
\generateur{ukwattaI}
\generateur{stamatikosJ}
\generateur{grupeJ}

\begin{table}
   \caption{Bursts with detected optical peaks.}
   
\begin{center}
\begin{tabular}{ccccccc}
  \hline
  \hline
  GRB & z  & $\egammaNorm$ & $\tpobs$ & $\tburst$  & References \\
  \hline
990123      & 1.60      &   436.52$\pm$60.31    &   47$\pm$10   &  63.3$\pm$0.3   &  \liangA , \lueC     \\
050820A   &  2.612   &  159.2$\pm$12.4   &    477$\pm$6  & 600$\pm$50   &  \lueC , \liangC     \\   
060418  &  1.489  &   48.6$\pm$10.6  &     170$\pm$5 &  52$\pm$1   &  \lueC , \liangC, \molinariG     \\
060605  &  3.78   &   2.8$\pm$0.5   &   590$\pm$45  & 19$\pm$1    &  \lueC , \liangC    \\
060607A &  3.082  &  23.4$\pm$1.5  &     179$\pm$3  & 100$\pm$5  &    \lueC , \liangC , \molinariG   \\ 
061007  &  1.261  &  421$\pm$41.9 &  77$\pm$1  &  75$\pm$5    &  \lueC , \liangC  \\ 
070318  &  0.836   &  1.3$\pm$0.3  &  507$\pm$46 &  63$\pm$5   &  \lueC , \liangC    \\
070419A &  0.97  &   0.2$\pm$0.02   &  765$\pm$30  & 112$\pm$2   &  \lueC , \liangC  \\
071010B  & 0.947  &  1.7$\pm$0.9   &  287$\pm$145 &  35.74$\pm$0.5   &  \lueC , \liangC \\ 
071031  &  2.692  &  3.9$\pm$0.6  &    1213$\pm$2 & 180$\pm$10   &  \stamatikosH, \liangC \\
080603A &  1.68742  &  2.2$\pm$0.8   &   1600$\pm$400  & 150$\pm$10   &  \martinI, \guidorziB   \\
080710  &  0.845  &  0.8$\pm$0.4   &   1934$\pm$46 & 120$\pm$17   &  \lueC , \liangC \\
080810  &  3.35   &  30$\pm$20    &   117$\pm$2  & 108$\pm$5  &  \lueC , \liangC \\ 
081008   & 1.967  &  2.8$\pm$0.5  &    163$\pm$2  & 185$\pm$39  & \yuanA, \liangC \\
081203A &  2.1   &   17$\pm$4    &   295$\pm$2 &  294$\pm$71  &  \ukwattaI, \liangC \\
090313  &  3.375  &  4.6$\pm$0.5  &    1315$\pm$109 & 78$\pm$19  &  \lueC , \liangC   \\  
090812  &  2.452  &  45.9$\pm$6  &   71$\pm$8  &  70$\pm$5  &  \stamatikosJ, \liangC  \\  
091029  &  2.752  &  7.4$\pm$074  &    328$\pm$50 &  39.2$\pm$5  & \grupeJ, \lueC \\
100906A &  1.727  &  33.4$\pm$3   &  101$\pm$4 &  114.4$\pm$1.6  &  \lueC , \liangC \\ 
110205A &  2.22  &   56$\pm$6   &    948$\pm$3 &  257$\pm$25  &  \lueC , \liangC \\
  \hline \\
\end{tabular}
\label{tab_detec}
\end{center}
References: 
\reftab{\liangA}{liang_2010};
\reftab{\lueC}{lue_2012};
\reftab{\liangC}{liang_2012};
\reftab{\molinariG}{molinari_2007};
\reftab{\stamatikosH}{stamatikos_2007};
\reftab{\martinI}{martin_2008};
\reftab{\guidorziB}{guidorzi_2011};
\reftab{\yuanA}{yuan_2010};
\reftab{\ukwattaI}{ukwatta_2008};
\reftab{\stamatikosJ}{stamatikos_2009};
\reftab{\grupeJ}{grupe_2009}
\end{table}

\subsection{Bursts with upper limits on $\tpz$}
\label{subsect_upperlim}

\begin{figure}
\begin{center}
\begin{tabular}{cc}
\includegraphics[width=0.43\textwidth]{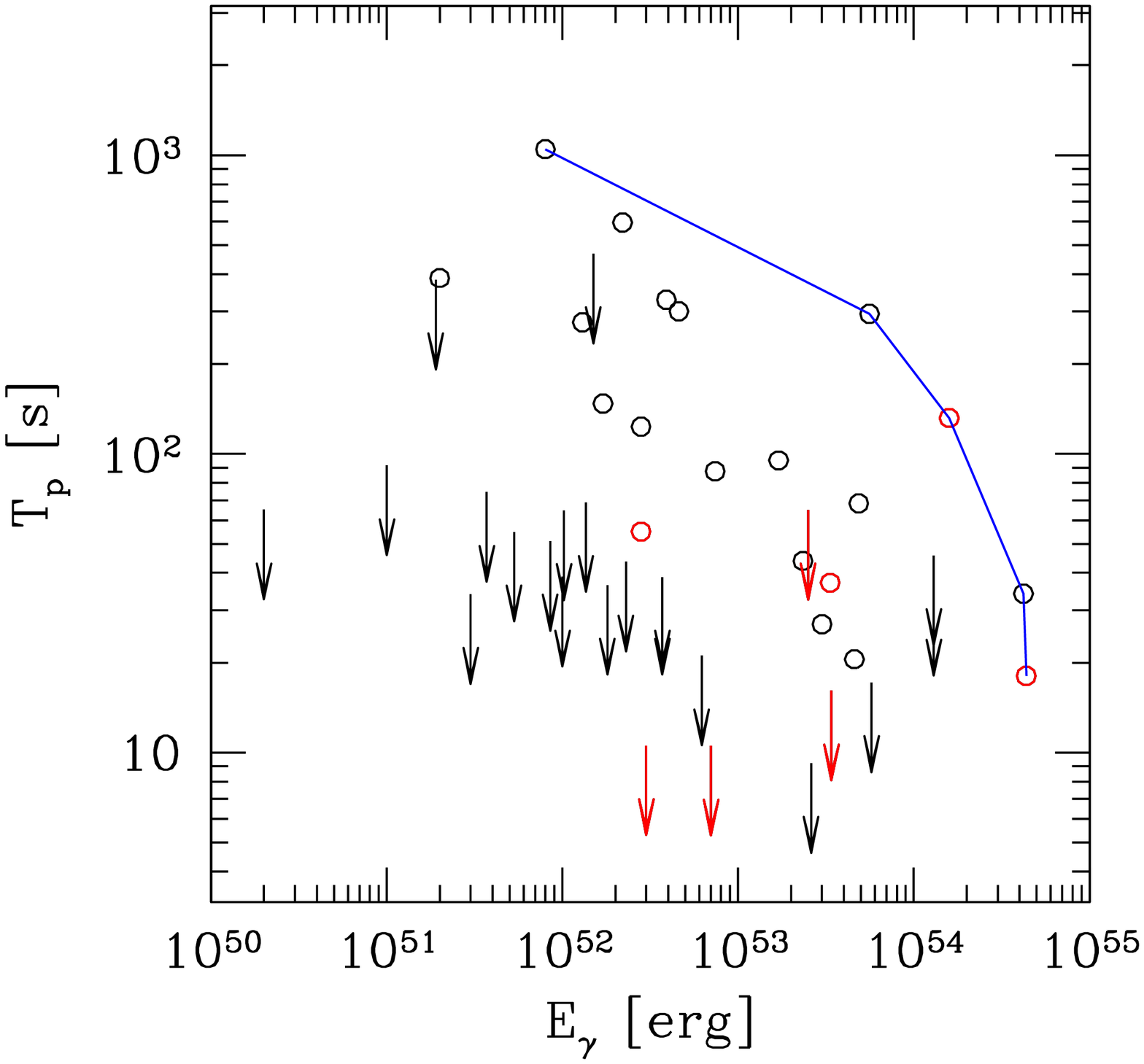} & 
\includegraphics[width=0.43\textwidth]{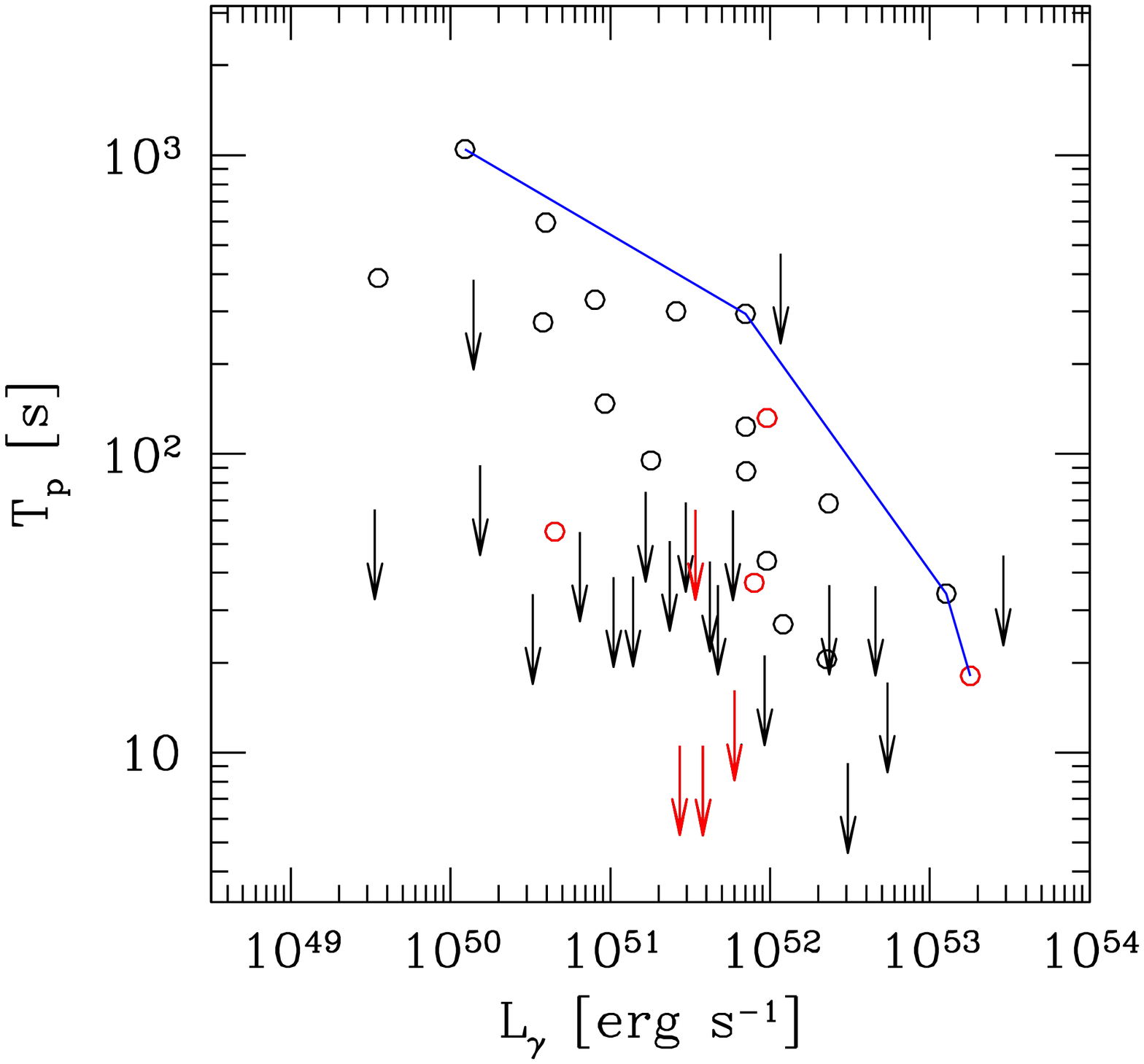}
\end{tabular}
\end{center}
\caption{Bursts on $\tpz-\egamma$ and $\tpz-\lgamma$ planes. 
\referee{Circles} represent bursts with detected $\tpz$, and arrows show upper limits.
Bursts with $\tpz<\tgamma$ are shown by red \referee{circles} and arrows.
The blue line shows the observed boundary of the burst population.
}
\label{fig_obs}
\end{figure}

For some bursts the peak is not observed because observations start too late.
These bursts are also useful for our purposes, as some of them give strong upper-limits on $\tpz$.
A sample of such GRBs is listed in \tab{tab_limits} and the corresponding upper limits are shown in \fig{fig_obs}. 
The number of available strong limits (24) is comparable to the number of peak detections. 
The limits are robust.
For many bursts in the sample, the optical decay was already well established when the observations started, without any evidence for an increasing decay index.
This suggests that the peak was reached well before the beginning of observations.

The obtained limits on $\tpz$ are never below a few tens of seconds, which reflects the typical delay in response of robotic optical telescopes to alerts from $\gamma$-ray telescopes.\footnote{Due to a fortunate chain of events, optical observations of GRB 080319B started \textit{before} the $\gamma$-ray trigger \citep{racusin_2008}. 
However in this special case the rise of the optical afterglow is hidden by the bright prompt optical component.}
Note also that the limits tend to be less constraining for weak (low $\egamma$) bursts. 
%
%
It is easier to obtain strong limits for bright bursts for a few reasons:
they are easily localized by $\gamma$-ray telescopes;
they have brighter afterglows \citep{gehrels_2008};
and they have higher redshifts which move $\tpz$ to a later $ \tpobs = (1+z) \tpz$

While it is easy to miss an early optical peak, $\tpz < 100$ s, 
we are not aware of any selection effects that could lead to preferential non-detection of late peaks $\tpz \sim 10^2 - 10^3$ s. 
In this range, the data should represent the true distribution of $\tpz$.
The data presented in \fig{fig_obs} may be summarized as follows: 
\textit{there is no intrinsic correlation between $\tpz$ and $\lgamma$ (or $\egamma$).
Instead we observe a lack of bright bursts with late afterglow peaks.}
For a given $\lgamma$, there appears to exist a maximum peak time $T_{\rm p, max} (\lgamma)$ which corresponds to the blue boundary in \fig{fig_obs}. 
A crude approximation to this boundary is given by $T_{\rm p, max}(\lgamma) \sim 200 \ \lgammaNorm^{-3/5} \ \mathrm{s}$.

\setcounter{refcompteur}{1}
\generateur{crewD}
\generateur{liD}
\generateur{pandeyD}
\generateur{wiersemaI}
\generateur{cusumanoG}
\generateur{quimbyG}
\generateur{sakamotoF}
\generateur{fynboF}
\generateur{rykoffF}
\generateur{blustinG}
\generateur{sollermanH}
\generateur{satoF}
\generateur{fugazzaF}
\generateur{zaninoniD}
\generateur{hunsbergerF} 
\generateur{crewF}
\generateur{krimmF}
\generateur{rykoffJ} 
\generateur{golenetskiiF}
\generateur{butlerG}
\generateur{krimmFb}
\generateur{cummingsG} 
\generateur{bloomG}
\generateur{depasqualeG}
\generateur{covinoA}
\generateur{dengJ}
\generateur{perleyI}
\generateur{huangC}
\generateur{racusinI}
\generateur{filgasB}
\generateur{jelinekI} 
\generateur{landsmanI} 
\generateur{oksanenI}
\generateur{cucchiaraI} 
\generateur{guidorziI}
\generateur{starlingJ}
\generateur{markwardtI}
\generateur{wrenI} 
\generateur{bergerI}
\generateur{gendreA}
\generateur{pageB}
\generateur{wiersemaC}
\generateur{ueharaC}

\begin{table}[h]
\caption{Bursts with upper limits on $\tpz$.}
\begin{center}
\begin{tabular}{cccccl}
  \hline
  \hline
  GRB & z  & $\egammaNorm$ & $\tpobs$ & $\tburst$ & References \\
  \hline
021211  &  1.004  &   1.02$\pm$0.1  &   $<130$ &  3.5$\pm$0.5 & \crewD, \liD, \pandeyD \\
040924  &  0.858  &  1.5$\pm$0.5     & $< 870$   &  2.39$\pm$0.24 & \wiersemaI \\
050319  &  3.24  &   3.7$\pm$1  &    $<164$  &   149.6$\pm$0.7 & \cusumanoG, \quimbyG  \\
050401  &  2.9  &   26$\pm$1  &    $<36$  &   33$\pm$2 & \sakamotoF, \fynboF, \rykoffF  \\
050525A &   0.606  &  2.3$\pm$1  &   $<70$   &   8.8$\pm$0.5 & \blustinG   \\
050824  &  0.828  &   0.19$\pm$0.05 &   $<700$  &    25$\pm$1 & \sollermanH  \\
050908  &  3.35  &   1.36$\pm$0.1 &   $<300$  &    20$\pm$2 & \satoF, \fugazzaF,\zaninoniD  \\
050922C &  2.17  &    3.7$\pm$1  &   $<116$ &  5$\pm$1 & \hunsbergerF, \crewF, \krimmF \\
051109A & 2.346  &   3$\pm$1   &    $<35$  &    37$\pm$5 & \rykoffJ, \golenetskiiF \\
051111  &  1.55   &  7$\pm$1  &  $<27$   &   47$\pm$1 & \butlerG, \krimmFb \\ 
060512A &  0.4428 &  0.02$\pm$0.005  &   $<94$  &    8.6$\pm$2 & \cummingsG, \bloomG, \depasqualeG  \\
060908  &  1.884  &  6.2$\pm$0.7   &  $<61$  &  19.3$\pm$0.3 & \covinoA \\
060912  &  0.937  &  0.85$\pm$0.15  &   $<99$   &   7$\pm$1 & \dengJ \\
071003  &  1.60435   &  34$\pm$4   &  $<42$    &  148$\pm$1 & \perleyI \\
071112C &  0.823  &  0.53$\pm$0.1   &   $<100$   &  15$\pm$12 & \huangC \\
080319B &  0.937  &  130$\pm$10    &  $<70$  &    55$\pm$5 & \racusinI \\
080413B  & 1.1   &   1.8$\pm$0.5  &      $<77$ &     8.0$\pm$1 & \filgasB \\
080430  &  0.767  &  0.3$\pm$0.1   &  $<60$   &   16.2$\pm$2.4 & \jelinekI, \landsmanI, \oksanenI, \cucchiaraI, \guidorziI \\
080721  &  2.591  &  130$\pm$10   &  $<164$   &   16.2$\pm$4.5 & \starlingJ  \\
081007  &  0.5295 &  0.1$\pm$0.02  &    $<140$   &   10.0$\pm$4.5 & \markwardtI, \wrenI, \bergerI \\
090102  &  1.547  &  57.5$\pm$5    &   $<44$   &   27$\pm$2.2 & \gendreA \\
090618  &  0.54  &  25.$\pm$1    &   $<100$   &   113$\pm$1 & \pageB \\
091018  &  0.971  &  0.37$\pm$0.1   &  $<147$   &  4.4$\pm$0.6 & \wiersemaC \\
091208B &  1.063  &  1$\pm$0.2  &   $<80$   &   14.9$\pm$3.7 & \ueharaC \\
  \hline
\end{tabular}
\label{tab_limits}
\end{center}
References: 
\reftab{\crewD}{crew_2003};
\reftab{\liD}{li_2003};
\reftab{\pandeyD}{pandey_2003};
\reftab{\wiersemaI}{wiersema_2008};
\reftab{\cusumanoG}{cusumano_2006};
\reftab{\quimbyG}{quimby_2006};
\reftab{\sakamotoF}{sakamoto_2005b};
\reftab{\fynboF}{fynbo_2005};
\reftab{\rykoffF}{rykoff_2005b};
\reftab{\blustinG}{blustin_2006};
\reftab{\sollermanH}{sollerman_2007};
\reftab{\satoF}{sato_2005};
\reftab{\fugazzaF}{fugazza_2005};
\reftab{\zaninoniD}{zaninoni_2013};
\reftab{\hunsbergerF}{hunsberger_2005};
\reftab{\crewF}{crew_2005};
\reftab{\krimmF}{krimm_2005};
\reftab{\rykoffJ}{rykoff_2009};
\reftab{\golenetskiiF}{golenetskii_2005};
\reftab{\butlerG}{butler_2006};
\reftab{\krimmFb}{krimm_2005b};
\reftab{\cummingsG}{cummings_2006};
\reftab{\bloomG}{bloom_2006};
\reftab{\depasqualeG}{depasquale_2006};
\reftab{\covinoA}{covino_2010};
\reftab{\dengJ}{deng_2009};
\reftab{\perleyI}{perley_2008};
\reftab{\huangC}{huang_2012};
\reftab{\racusinI}{racusin_2008};
\reftab{\filgasB}{filgas_2011};
\reftab{\jelinekI}{jelinek_2008};
\reftab{\landsmanI}{landsman_2008};
\reftab{\oksanenI}{oksanen_2008};
\reftab{\cucchiaraI}{cucchiara_2008};
\reftab{\guidorziI}{guidorzi_2008};
\reftab{\starlingJ}{starling_2009};
\reftab{\markwardtI}{markwardt_2008};
\reftab{\wrenI}{wren_2008};
\reftab{\bergerI}{berger_2008};
\reftab{\gendreA}{gendre_2010};
\reftab{\pageB}{page_2011};
\reftab{\wiersemaC}{wiersema_2012};
\reftab{\ueharaC}{uehara_2012}
\end{table}

\begin{figure}[h]
\begin{center}
\begin{tabular}{c}
\includegraphics[width=0.5\textwidth]{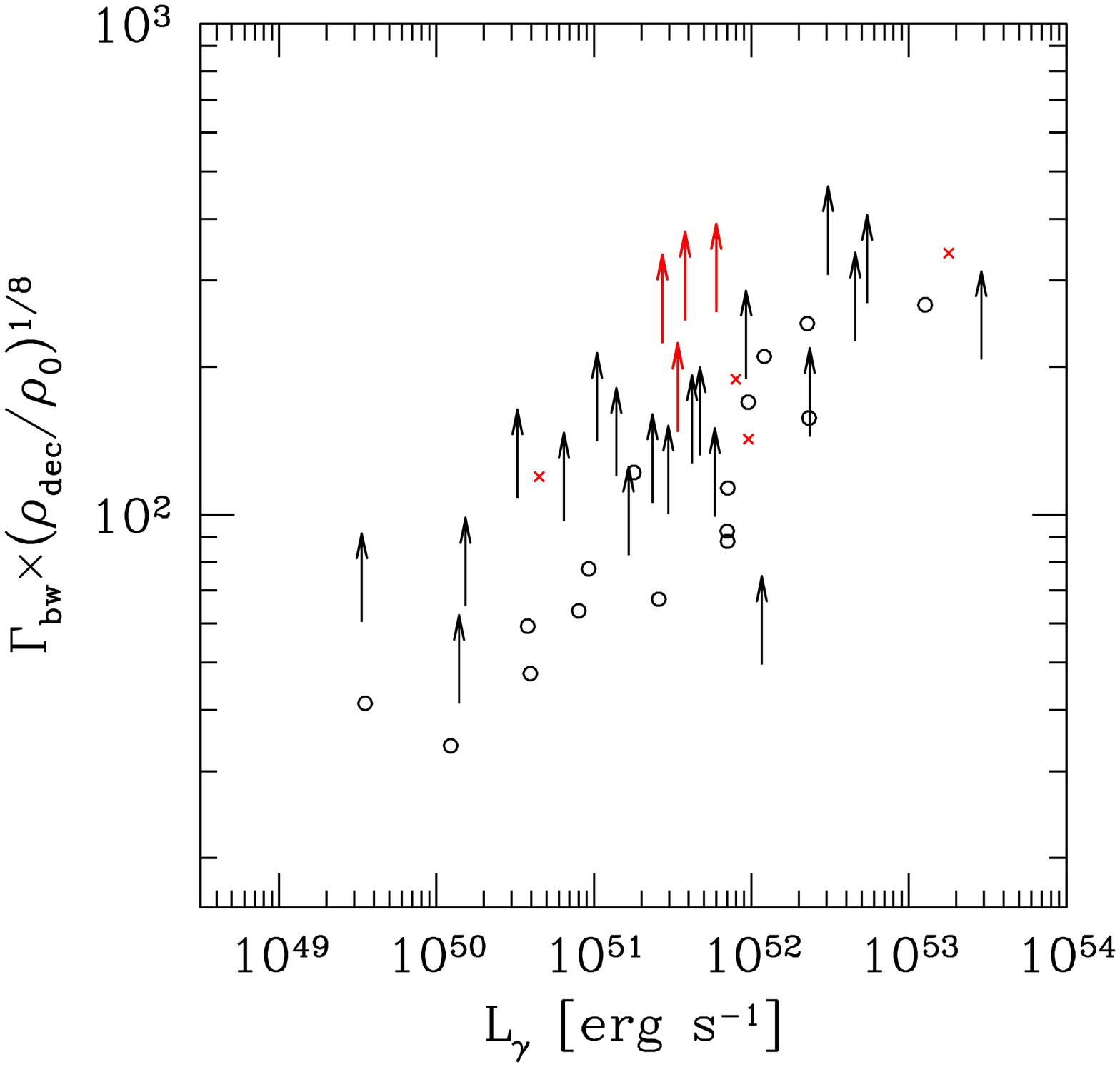} 
\end{tabular}
\end{center}
\caption{Estimated Lorentz factor of the GRB blast wave, $\lfbw$, at the deceleration radius. 
The product $\lfbw  \left. \rhodec \right. ^{1/8}$ is estimated using \eq{eq_lf_estimate} (with $\eta=0.5$) 
and shown versus the burst luminosity $\lgamma$ (\eq{eq_lum}).
The ambient density at the deceleration radius, $\rhodec$, is normalized to $\rho_0 /m_{\rm p} = 1 \ \mathrm{cm}^{-3}$.
Bursts with $\tpz<\tgamma$ are highlighted in red;
for these bursts the ejecta Lorentz factor $\lfej$ can be substantially higher than $\lfbw$ (see text).}
\label{fig_inter}
\end{figure}


\begin{figure}[h]
\begin{center}
\begin{tabular}{c}
\includegraphics[width=0.47\textwidth]{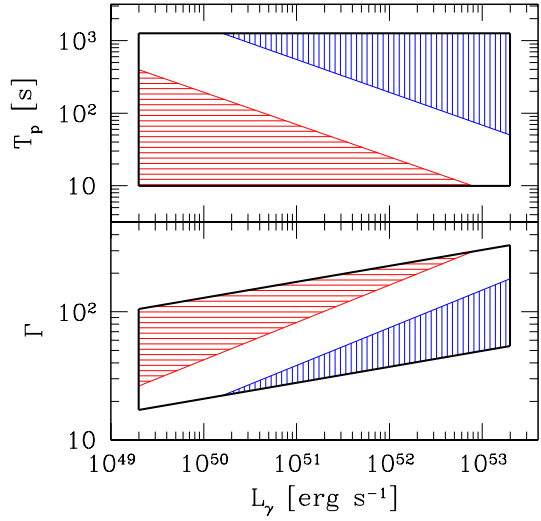} 
\end{tabular}
\end{center}
\caption{Effects of the $\tpz \rightarrow \lfej$ transformation.
This schematic figure illustrates how the GRB population presented in \fig{fig_obs} ($\tpz-\lgamma$ plane) transforms into the $\lfej-\lgamma$ plane.
The hatched and colored areas are indicated to better visualize how different regions on the $\tpz-\lgamma$ plane transform to the $\lfej-\lgamma$ plane.
The red (horizontally hatched) region is where strong selection effects are expected to suppress the observed population.
Combined with the real lack of bursts in the blue (vertically hatched) region, this leads to a spurious correlation between $\tpz$ and $\lgamma$.
The corresponding spurious $\lfej-\lgamma$ correlation is enhanced by the deformation of the population in the $\lfej-\lgamma$ coordinates -- 
the black rectangular in the upper panel is transformed into a ``parallelogram'' in the lower panel.
}
\label{fig_cartoon}
\end{figure}


\begin{figure}[h]
\begin{center}
\begin{tabular}{c}
\includegraphics[width=0.47\textwidth]{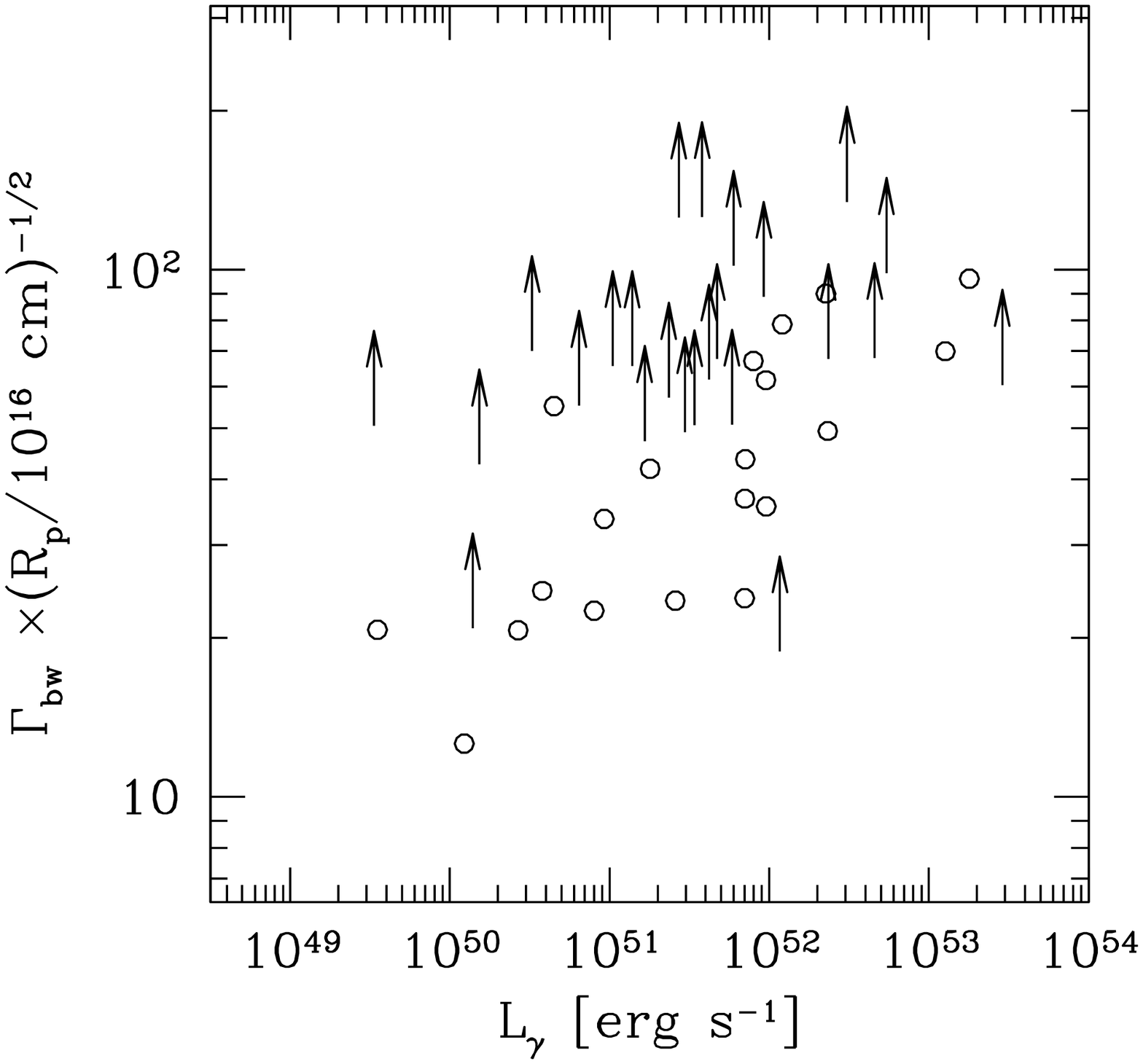} 
\end{tabular}
\end{center}
\caption{Estimated Lorentz factor of the GRB blast wave $\lfbw = (\rp/2c\tpz)^{1/2}$ for an arbitrarily fixed afterglow peak radius $\rp$. 
The product $\lfbw \rp^{-1/2}$ shown in this figure is proportional to $\tpz^{-1/2}$, 
so the diagram is a simple transformation of the $\tpz - \lgamma$ diagram shown in \fig{fig_obs}.
}
\label{fig_rp}
\end{figure}


\section{Estimates for $\lfej$}
\label{lf_dist}

The GRB afterglow is likely emitted by the blast wave resulting from the interaction of the relativistic ejecta with the ambient medium. 
The blast wave involves two shocks: the forward shock sweeping the external medium and the reverse shock propagating back into the relativistic ejecta. 
As discussed in \sect{sect_intro}, it is reasonable to assume that the afterglow peaks at the deceleration time $\tdec$
when most of the ejecta energy has been transmitted to the blast-wave through the reverse shock. 
This happens at the ``deceleration radius,''
\begin{equation} 
\label{eq_rdec}
\rdec = \left( \frac{3-s}{4 \pi c^2} \frac{E_{\rm  ej}}{\lfbw^2 \rhodec} \right)^{\frac{1}{3}} \, ,
\end{equation}
where $E_{\rm  ej}$ is the energy of the ejecta and 
$\lfbw$ is the Lorentz factor of the blast wave at $\rdec$; 
$\rhodec$ is the external density at $\rdec$, and $s$ describes the slope of the external
density profile, $s = \mathrm{d} \ln \rho / \mathrm{d} \ln R$.
A uniform medium is described by $s=0$ and a wind medium by $s=2$.
We assume that the afterglow peaks at the deceleration time,
\begin{equation} 
\label{eqn_tpEtdec}
\tpz \simeq \tdec \simeq \frac{\rdec}{2 \lfbw^2 c} \, .
\end{equation}
Then the measured $\tpz$ provides an estimate for the blast wave Lorentz factor,
\begin{equation} 
\label{eq_lf_estimate}
\lfbw = \left[ \frac{3-s}{32 \pi c^5} \frac{1-\eta}{\eta} \frac{\egamma}{\rhodec \tpz^3}  \right]^{\frac{1}{8}} \, ,
\end{equation}
where $\eta$ is the fraction of the initial energy of the GRB ejecta that is converted into prompt radiation.

Two aspects of \eq{eq_lf_estimate} should be noted: 
(1) the estimate depends on the poorly known ambient density as $\left[ \rhodec / (3-s) \right]^{-1/8}$ and the prompt efficiency as $\left[ \eta / (\eta-1) \right]^{-1/8}$.
(2) The estimate gives the Lorentz factor of the \textit{blast wave}, $\lfbw$, not the
ejecta Lorentz factor $\lfej$.

We expect $\lfej \simeq \lfbw$ if $\tpz \gg \tgamma$. 
Indeed, $\tpz$ is associated with the time it takes the ejecta to transfer most 
of its energy to 
the blast wave, i.e. the time it takes the reverse shock to cross the main,
most energetic part of the ejecta of thickness $\Delta \simeq c \tgamma$.
The crossing time $\Delta / v_{\rm rs}$ is long and gives $\tpz \sim (c/v_{\rm rs})\tgamma \gg \tgamma$ if the reverse shock is non relativistic, $v_{\rm rs}\ll c$,
which is equivalent to $\lfej \simeq \lfbw$. In this case \eq{eq_lf_estimate} effectively
gives an estimate of the mean Lorentz factor $\lfej$ of the ejecta.
In contrast if $\tpz \la \tgamma$, the reverse shock may be highly relativistic. 
Then $\lfbw$ is significantly smaller than $\lfej$ and \eq{eq_lf_estimate} significantly underestimates $\lfej$. 
GRBs with $\tpz \la \tgamma$ are highlighted in red in \fig{fig_obs}.

Note that $\tgamma$ that we use as a measure of the GRB duration may overestimate the duration of the main part of the GRB 
if the burst has a temporally extended tail of relatively weak emission.
For such bursts $\tdec < \tgamma$ is possible.
A better estimate for $\tgamma$ would give $\tdec \simeq \tgamma$, so that $\tpz$ is not smaller than $\tgamma$. 
In agreement with theoretical expectations, we found no burst in our sample 
where the optical afterglow peaks before the main part of the GRB emission has been received.

\eq{eq_lf_estimate} assumes a static external medium and neglects the fact that the prompt GRB radiation exerts pressure and accelerates the medium ahead of the blast wave \citep{thompson_2000, beloborodov_2002}.
This pre-acceleration is strong (relativistic) up to the radius,
\begin{equation}
R_{\rm acc} = 2 \times 10^{15} (\egammaNorm)^{1/2} \ \mathrm{cm} \, .
\end{equation}
If $R_{\rm acc}$ exceeds $\rdec$ given by \eq{eq_lf_estimate}, the true deceleration radius is increased, and the dissipation rate peaks at
\begin{equation}
\tpz =  \frac{R_{\rm acc}}{2 \lfbw^2 c} \simeq 4 (\egammaNorm)^{1/2} \left(\frac{\lfbw}{100}\right)^{-2} \ \mathrm{s}   \, .
\end{equation}
We found $R_{\rm acc} < \rdec$ for all bursts in our GRB sample 
as long as $\rhodec / m_{\rm p} \la 10^3 \ \mathrm{cm}^{-3}$. 
For most bursts in our sample, $\rhodec / m_{\rm p} \ga  10^3 \ \mathrm{cm}^{-3}$ would imply low values for $\lfej$, which would contradict the constraint from the prompt emission (see discussion in \sect{par_ambient}).
Therefore we will assume $\rhodec / m_{\rm p} \la  10^3 \ \mathrm{cm}^{-3}$ and neglect the pre-acceleration effect.

Using \eq{eq_lf_estimate} we estimated $\lfej$ for each burst in the sample (\fig{fig_inter}).
We fixed $\eta=0.5$ and $s=2$ in our numerical estimates; the uncertainty in their exact values weakly affects the results.
If only the peak ``detections'' are considered, \fig{fig_inter} would suggest that there is a correlation between $\lfej$ and $\lgamma$. 
However, the numerous lower limits show that the upper part of the diagram must be broadly populated by GRBs.
Also note that four detections of $\tpz$ correspond to the relativistic reverse shock regime and give only lower limits on $\lfej$.
We conclude that there is no evidence for
a correlation between $\lgamma$ and  $\lfej$. 
However the lack of bright bursts with low $\lfej$ appears to be a robust feature.

\fig{fig_cartoon} illustrates how the GRB sample is transformed from 
$\lgamma-\tpz$ plane to $\lgamma-\lfbw$ plane, 
following the relation between $\lfbw$ and $\tpz$, $\lfbw \propto \egamma^{1/8} \tpz^{-3/8}$ (\eq{eq_lf_estimate}).
This transformation compresses the sample along the $\lfbw$ axis and induces 
a dependence of $\lfbw$ on $\lgamma$ with a positive slope of $1/8$.
In combination with the selection effect that suppresses short-$\tpz$ and low-$\lgamma$
 bursts in the sample, this
enhances the 
spurious correlation between $\lgamma$ and  $\lfej$.

As mentioned in \sect{sect_intro}, the key assumption that the optical afterglow peaks at the deceleration radius may not be reliable.
Therefore, it is useful to consider a more general model where the afterglow peaks at a radius $R_{\rm p}$ related to $\lfej$ and $\lgamma$ by
\begin{equation}
\rp \propto \lfej^{\alpha} \lgamma^{\beta} \, .
\end{equation}
The simplest model with $\rhodec = \mathrm{const}$ corresponds to $\alpha = -2/3$ and $\beta = 1/3$ (see \eq{eq_rdec}).
For other values of $\alpha$, $\beta$, the observed upper boundary $\tpz^{\rm max} \propto \lgamma^{\lambda}$ (where $\lambda \sim -3/5$; see \fig{fig_obs}) in the $\lgamma - \tpz$ diagram still transforms into a lower-boundary $\lfej_{\rm min} \propto \lgamma^{(\lambda-\beta)/(\alpha-2)}$ in the $\lgamma - \lfej$ diagram.
As an illustration,  \fig{fig_rp} shows $\lfej$ estimated assuming $\alpha = \beta = 0$  (i.e. $\rp = \mathrm{const}$). 
The results are similar to those in \fig{fig_inter}.
The conclusion that $\lfej_{\rm min}$ grows with $\lgamma$ holds as long as $(\lambda-\beta)/(\alpha-2) > 0$.
Violation of this condition would require $\alpha$ and $\beta$ that are significantly different
from $\alpha = -2/3$ and $\beta = 1/3$.
If $\alpha \simeq 2$ then $\tpz$ weakly depends on $\lfej$, and cannot be used to estimate $\lfej$.


\section{Lack of bright bursts with a large $\tpz$: a result of adiabatic cooling?}
\label{opacity_effect}

\subsection{Adiabatic cooling below the photosphere}
\label{pres_cooling}

The photospheric radius $\rph$ (the characterisitic radius where the explosion ejecta becomes transparent to Thomson scattering) is given by
\begin{equation}
\label{ph_radius}
\rph \approx 3 \times 10^{12} \kappa_{0.2} L_{52} \lfej_{2}^{-3} \ \mathrm{cm} \, ,
\end{equation}
where $L$ is the isotropic power of the outflow, and
$\kappa$ is the Thomson opacity (in units of $0.2 \ \mathrm{cm^2 g^{-1}}$);
$\kappa$ may be significantly increased by pair creation.
For the most powerful explosions ($\lgamma \sim 10^{53}-10^{54} \ \mathrm{erg \ s^{-1}}$) with relatively low Lorentz factors ($\lfej \la 10^2$) the photospheric radius is exceptionally large.
If the GRB
radiation is produced at a smaller radius $\rdiss$ (where internal dissipation peaks) the burst may be buried by the large optical depth,
since it implies strong adiabatic cooling of the radiation trapped in the expanding ejecta.

To illustrate this possibility, suppose that the GRB emission is generated before the ejecta reaches a radius 
\begin{equation}
\label{eq_rdiss}
\rdiss = \lfej^2 R_{0} \, , 
\end{equation}
where $R_{0}$ is a fixed constant.
The scaling of the cutoff radius $\rdiss \propto \lfej^2$ is expected for mechanisms that dissipate the energy of internal motions or magnetic energy.
Combining Equations (\ref{ph_radius}) and (\ref{eq_rdiss}), one finds the optical depth at $\rdiss$,
\begin{equation}
\tau_{\rm diss} = \frac{\rph}{\rdiss} \propto L \lfej^{-5} \, .
\end{equation}

If dissipation occurs far below the photosphere ($\tau_{\rm diss} \gg 1$), the resulting radiation released at the photosphere is 
adiabatically cooled by the factor
of $2\tau_{\rm diss}^{-2/3}$  \citep{beloborodov_2011}.
Then for given ejecta power
$L$ and dissipation efficiency at $\rdiss$ the observed
burst luminosity scales as 
\begin{equation}
\lgamma \propto \lfej^{10/3} \, .
\end{equation}
The slope of this relation is suggestively close
to the slope of $\lfej_{\rm min} - \lgamma$ relation seen in \fig{fig_inter}.

\subsection{Properties of adiabatically cooled bursts}
\label{subsect_prop}

Even though the dissipation mechanism at $r<\rdiss$ can be non-thermal, 
the produced radiation will be progressively thermalized during the subsequent 
adiabatic expansion between $\rdiss$ and $\rph$.
Let $T_{\rm e}$ be the electron temperature (measured in the ejecta frame), and $\tau_{\rm diss}>1$ be the Thomson optical depth at $\rdiss$.
Two different regimes can be distinguished:
(1)  For $\tau_{\rm diss} \la m_e c^2/3 k T_e \simeq 511 \left(3 k T_e / 1 \ \mathrm{keV}\right)^{-1} $, 
an exponential cutoff would form in the radiation spectrum at energy
\begin{equation}
\label{eqn_ecut}
\ecut \approx\frac{\lfej m_e c^2}{\tau_{\rm diss}} \approx 511\,\frac{\lfej}{\tau_{\rm diss}}  \ \mathrm{keV} \, .
\end{equation}
This cutoff is a result of significant Compton downscattering (recoil effect) at $E > \ecut$; the 
spectrum at $E < \ecut$ is weakly affected.
(2) For $\tau_{\rm diss} \ga m_e c^2/3 k T_e$, 
the spectrum is exponentially suppressed above $3kT_e\lfej$. 
The low-energy part of the spectrum is also affected by multiple Compton scattering ---
the spectral slope steepens as the photons tend to thermalize with electrons. 

Thus, one expects significant changes in the burst spectrum after strong adiabatic cooling.
Such unusual GRBs have been observed.
\citet{ghirlanda_2003} and \citet{ryde_2004} found that some bursts have spectra with very hard low energy indices and possibly exponential cutoffs. 
Similar quasi-thermal GRBs are found in both BATSE \citep{kaneko_2006} and 
Fermi Gamma Burst Monitor \citep{goldstein_2012} catalogs.
These bursts --- especially 
those with a low peak energy ---  may be generated by 
adiabatically cooled explosions with a large photospheric radius $\rph$. 

Another expected feature of bursts with small $\lfej$ and large $\rph$
is the suppression of variability on short timescales. In these bursts,
the minimum variability timescale $\dtobs \sim \rph / 2 \lfej^2 c$ can be as large as $10$~s 
(see \eq{eqn_dtmin} below), 
and their lightcurves are expected to be smooth.

While adiabatic cooling can significantly reduce the emitted GRB energy $\egamma$, the ejecta energy $E_{\rm ej}$ remains large. 
Thus, adiabatically cooled bursts are expected to have unproportionally 
bright afterglows.
They should lie in the upper part of the gamma-ray fluence/X-ray afterglow flux distribution,
which spreads over two orders of magnitude (see for example Fig. 2 in \citealt{gehrels_2008}).
As we show below, the majority of GRBs with strong adiabatic cooling avoid detection.
Then they become prime candidates for ``orphan'' afterglows (e.g. \citealt{huang_2002}).
These afterglows are expected to peak at late times $\tpz$, as they are generated by 
the low-$\lfej$ ejecta.

\subsection{Monte Carlo simulation of a GRB population}
\label{monte_carlo}

To illustrate how bright bursts with low Lorentz factors are depleted by adiabatic cooling, 
we produced a synthetic GRB population using a Monte Carlo simulation,
with the following assumptions: \\
(1) We adopted the GRB rate R$_{SF3}$ from \citet{porciani_2001}.
We assume that the rate
keeps increasing at $z \ga 2$ as suggested by observations (e.g. \citealt{daigne_2006, wanderman_2010, salvaterra_2012});
we cut it off at $z_{\rm max} \simeq 20$. \\
(2) The GRB luminosity function is assumed to follow a power-law distribution of index $-1.5$ as suggested by \citet{daigne_2006}, 
in a broad range of 
$10^{50} < \lgamma < 10^{54} \ \mathrm{erg \ s^{-1}}$. \\
(3) For each GRB, the spectrum of generated radiation at $\rdiss$
is assumed to be a broken power-law with a low energy index $\alpha = -1$ and a high energy index $\beta = -2.5$. 
The \referee{rest-frame} peak energy of the spectrum, $\ep$, is assumed to correlate with $\lgamma$;
we use the relation 
$\ep \simeq 300\, (\lgamma / 10^{52} \ \mathrm{erg \ s^{-1}})^{1/2} \ \mathrm{keV}$
with a scatter $\sigma_{\rm dex} = 0.3$
 (e.g. \referee{\citealt{wei_2003, yonetoku_2004, nava_2012}}).\\
(4) The radiation spectrum is injected with $\eta=0.5$ at $\rdiss = 6 \times 10^{12} \ (\lfej/100)^2 \ \mathrm{cm}$ 
(which would correspond to a variability timescale 
   $\Delta t_{\rm obs} = \rdiss / 2 \lfej^2c \simeq 10$~ms if $\rdiss>\rph$).\\
(5) The logarithm of Lorentz factor $\lfej$ is randomly chosen for each burst from
a uniform distribution in the range $1<\log \lfej<3$.\\

The adiabatic cooling effect is calculated as follows:\\
(6) The photospheric radius of each GRB is obtained from \eq{ph_radius}, with $\kappa_{0.2} = 1$.
If $\rdiss < 2^{-3/2} \rph$, the burst is cooled by a factor of $2 (\rph/\rdiss)^{-2/3}$ changing $\lgamma$ and $\ep$ from their initial values at $\rdiss$. 
The burst spectrum is changed as explained in \sect{subsect_prop}; the cutoff at $\ecut$ is approximated by a step function.\\
(7) The simulated GRB is assumed to be detected if its observed photon flux in the \textit{Swift} band $15-150$ keV is above the threshold of $0.2 \ \mathrm{ph \ cm^{-2} \ s^{-1}}$ \citep{band_2006}.\\

\begin{figure}[h]
\begin{center}
\begin{tabular}{cc}
\includegraphics[width=0.47\textwidth]{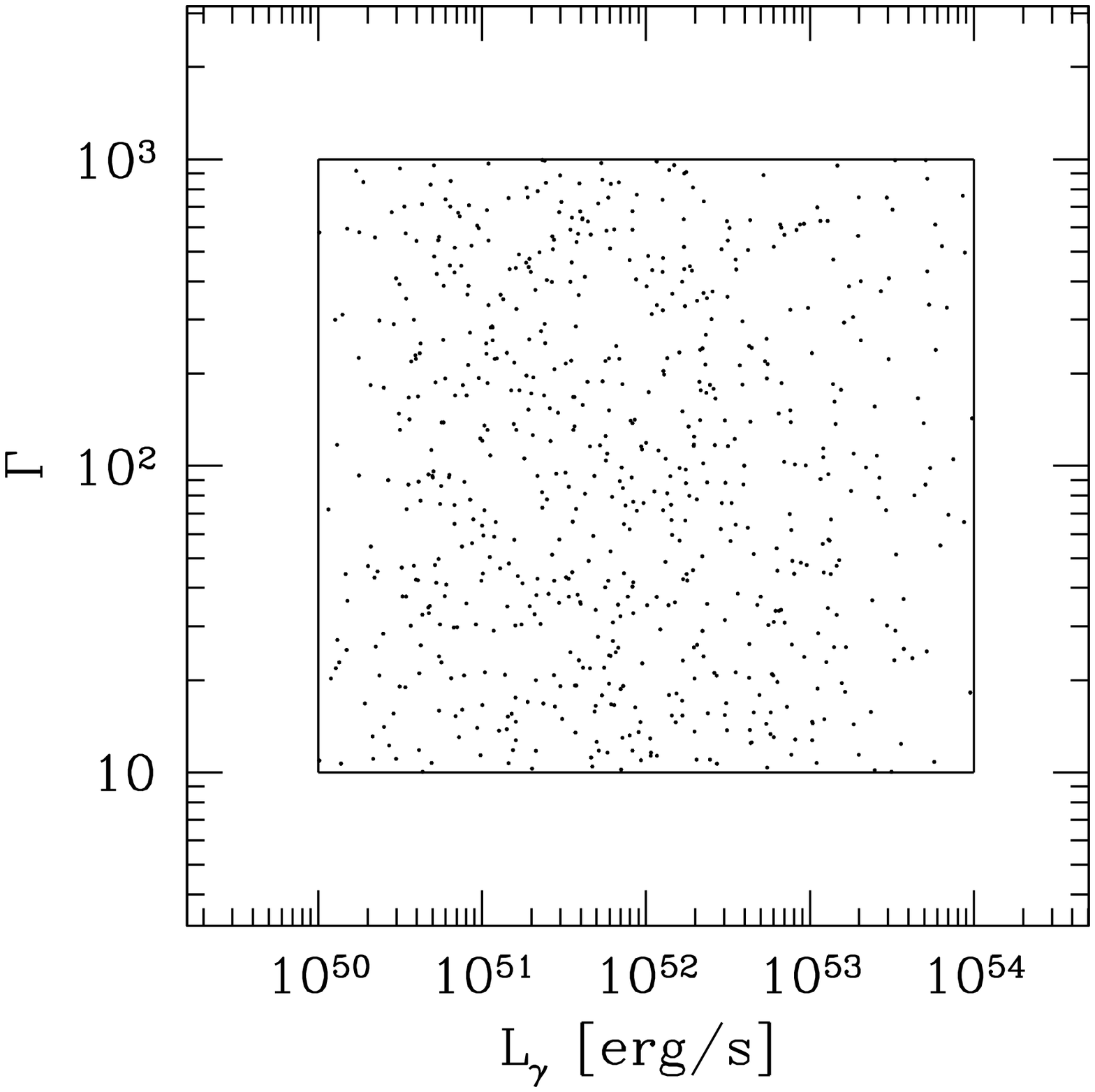} &
\includegraphics[width=0.47\textwidth]{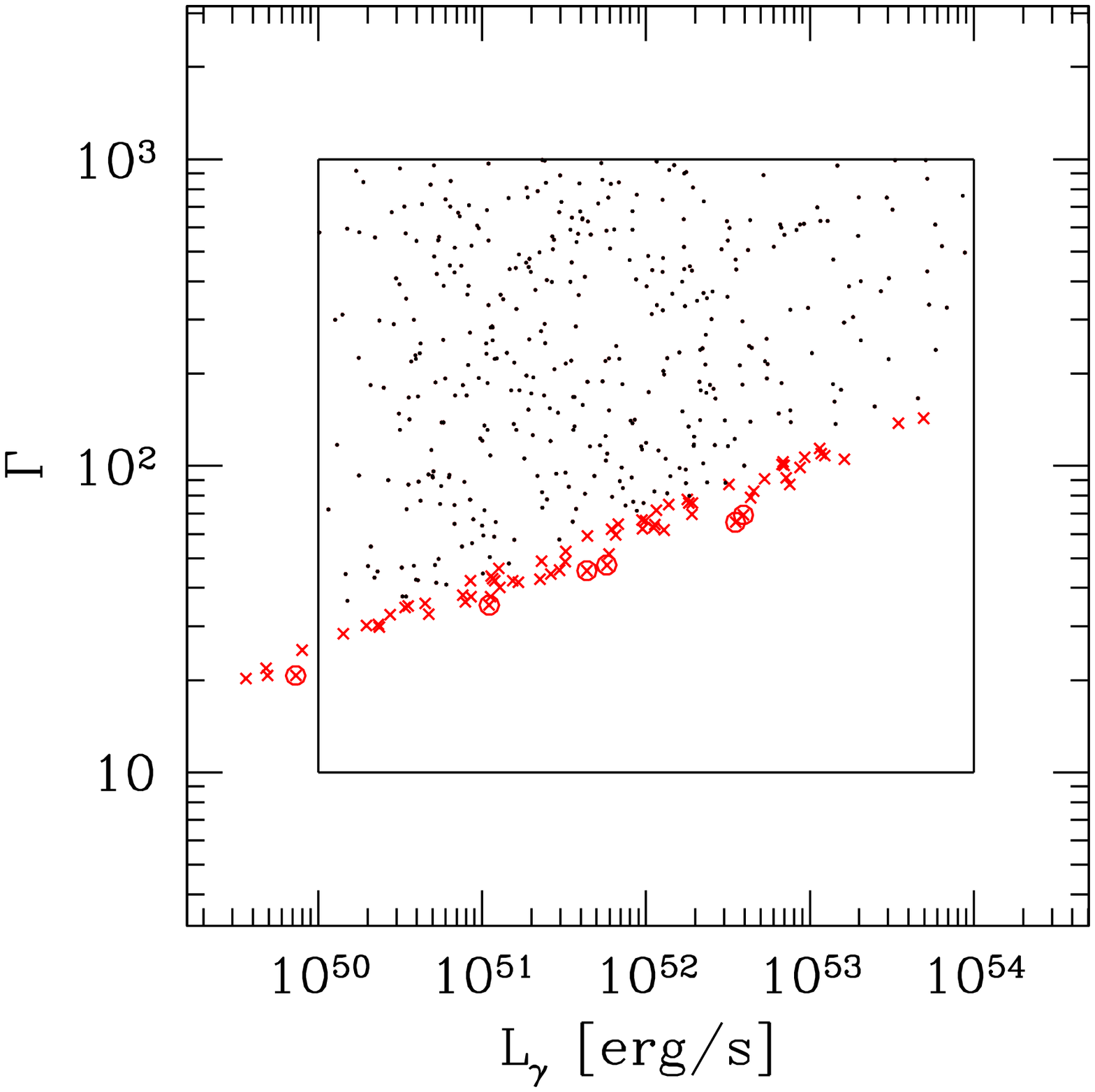}
\end{tabular}
\end{center}
\caption{Effect of adiabatic cooling on a GRB population in the $\lgamma-\lfej$ plane.
\textit{Left:} observed distribution of bursts when adiabatic cooling is not included. 
\textit{Right:} observed distribution of bursts when adiabatic cooling is included.
Red crosses represent bursts that have suffered cooling and remained detectable.
Many more cooled bursts became undetectable and disappeared from the diagram.
The circled red crosses show detected bursts with large cooling factors $f>10$ (1.3\% of detected bursts).
These bursts are expected to have special spectra of the prompt GRB emission (see text).
}
\label{fig_mc_lf}
\end{figure}

\begin{figure}[h]
\begin{center}
\begin{tabular}{cc}
\includegraphics[width=0.47\textwidth]{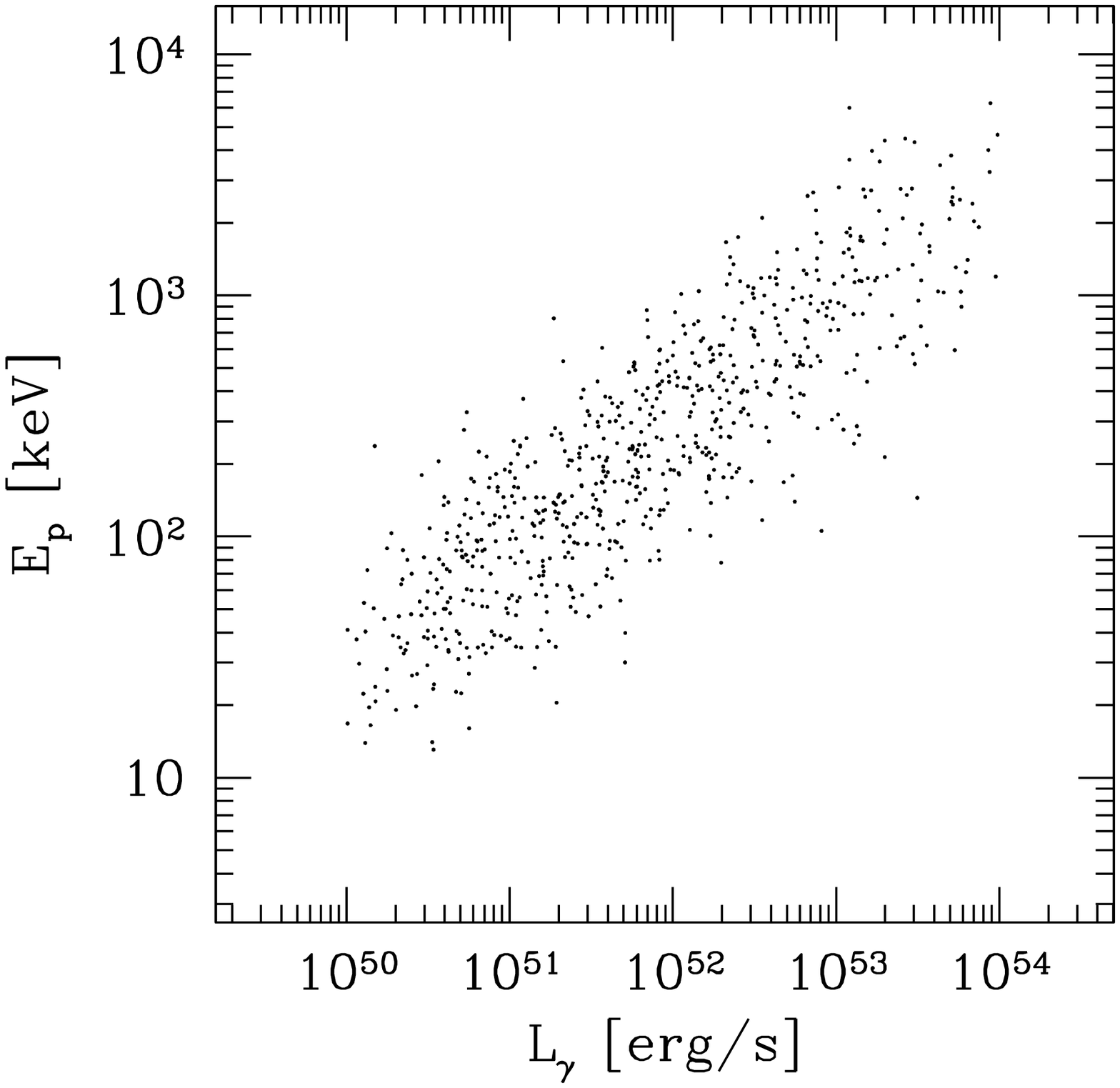} &
\includegraphics[width=0.47\textwidth]{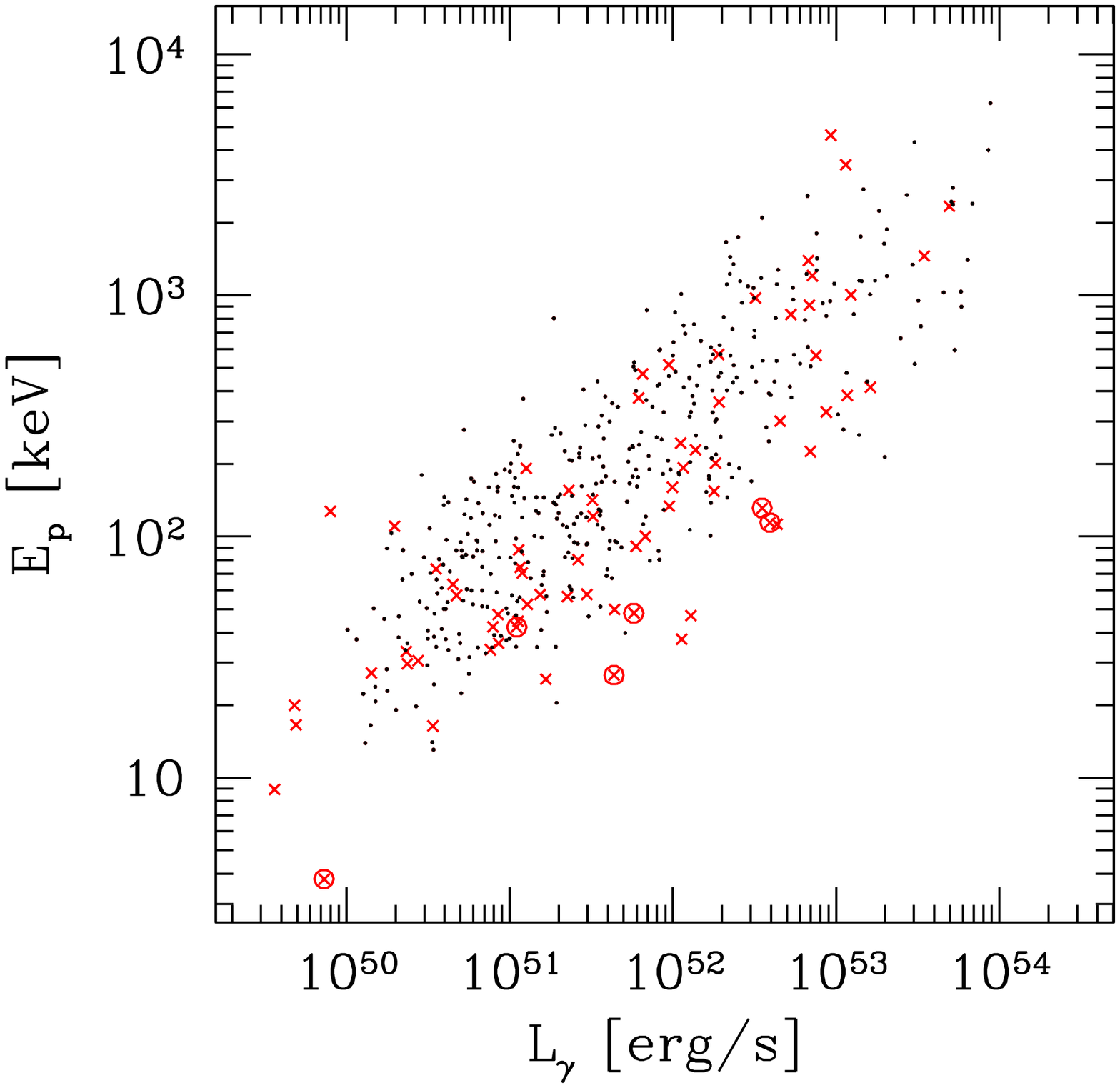}
\end{tabular}
\end{center}
\caption{Effect of adiabatic cooling on a GRB population in the $\lgamma-E_{\rm p}$ plane.
The simulation is the same as in \fig{fig_mc_lf}.
\textit{Left:} observed distribution of bursts when adiabatic cooling is not included. 
\textit{Right:} observed distribution of bursts when adiabatic cooling is included.
Red crosses represent bursts that have suffered cooling.
The circled red crosses show detected bursts with large cooling factors $f>10$ (1.3\% of detected bursts).
Strong adiabatic cooling can in principle create outliers of the initial $\ep - \lgamma$ correlation,
however most of them become undetectable and disappear from the diagram.
}
\label{fig_mc_ep}
\end{figure}

The results of our simulation are shown in \fig{fig_mc_lf}.
One can see that adiabatic cooling depletes the low-$\lfej$/high-$\lgamma$ 
corner of the $\lgamma - \lfej$ diagram.
The resulting distribution resembles the observed one in \fig{fig_inter}.
This simulation also allows one to estimate the impact of adiabatic cooling on the 
observed distribution of $\ep$ (\fig{fig_mc_ep}).
A burst that suffers adiabatic cooling is moved along a track $\ep \propto \lgamma$, which tends to create GRBs with $\ep$ below the original correlation.
However, the effect on the population of detected GRBs is weak, because cooled bursts become undetected (due to the reduced $\ep$ and the spectral cutoff at $\ecut$)
before they become outliers in the $\ep - \lgamma$ correlation.


\section{Discussion}
\label{conclusion}

\subsection{Adiabatic cooling}

If the peak time of optical afterglow, $\tpz$, is indeed a good proxy for the blast wave deceleration radius $\rdec$,
observations imply a lack of bright bursts with low Lorentz factors (\sect{lf_dist}).
We argued that this 
lack may be expected, as energetic explosions with low Lorentz factors 
should have unusually large photospheres $\rph \ga 10^{15}$~cm. 
Since the dissipation
mechanism generating radiation in these bursts is likely limited to smaller radii,
$\rdiss\ll\rph$, the burst is expected to suffer strong adiabatic cooling and become
undetectable (Section 4).

Most models of the prompt GRB emission place the emission source at radii smaller than $10^{15}$~cm, especially if $\lfej$ is small.
For example, dissipation of internal motions or magnetic energy in the ejecta 
is expected to end at a radius that scales as $\lfej^2$ and becomes smaller than $\rph$ at small $\lfej$. 
Then the trapped radiation is adiabatically cooled and the burst becomes 
inefficient. This argument is applicable to any dissipation mechanism 
generating the burst
--- e.g. collisionless shocks \citep{rees_1994, daigne_1998}, collisional heating
\citep{beloborodov_2010}, or magnetic reconnection \citep{spruit_2001}.
Note also that the neutron component of the jet, which can 
play a significant role in collisional dissipation, may not survive to $\rph$ in low-$\lfej$
bursts. The mean radius of neutron decay 
is $R_{\beta} \simeq 9 \times 10^{13} (\lfej/30)$~cm, and its ratio to the photospheric 
radius is given by
\begin{equation}
   \frac{R_{\beta}}{\rph} \simeq 7.5 \times 10^{-2} \left.\kappa_{0.2}\right.^{-1} 
\left.L_{54}\right.^{-1} \left( \frac{\lfej}{30} \right)^4 \ll 1 \, .
\end{equation}

The cooled bursts still produce energetic ejecta that can drive an energetic blast wave
in the external medium and generate bright afterglow emission. We argued that one
could observe ``orphan'' afterglows from such explosions, with undetected prompt GRBs,
even when the burst is observed ``on-axis,'' i.e. the relativistic jet is directed toward 
the observer. As this paper was completed, Palomar Transient Factory detected an 
event consistent with on-axis orphan afterglow \citep{cenko_2013}.
  
We also argued that the prompt emission of strongly cooled bursts can be occasionally detected. 
As discussed in Section~4, these bursts have special properties. 
They should have soft spectra resembling quasi-thermal emission and their light curves
should be smooth.

\subsection{GRB ambient medium: a low density wind?}
\label{par_ambient}


\begin{figure}[h]
\begin{center}
\begin{tabular}{c}
\includegraphics[width=0.47\textwidth]{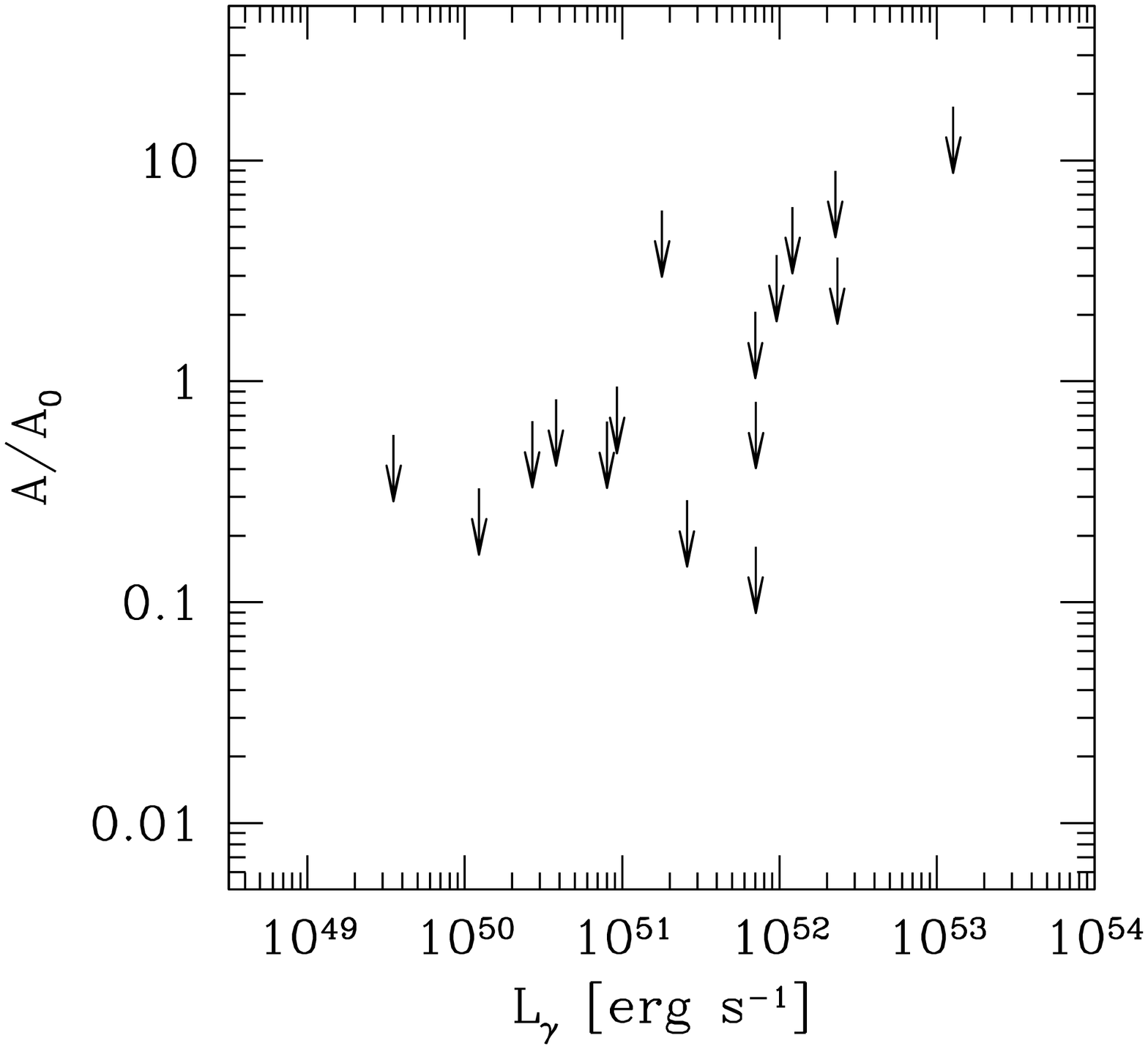} 
\end{tabular}
\end{center}
\caption{Constraints on the wind from the GRB progenitor.
The figure shows the derived upper limits on the wind density parameter $A=\dot{M}/4\pi w$ in units of $A_0 = 5 \times 10^{11} \ \mathrm{g \ cm^{-1}}$.
}
\label{fig_awind}
\end{figure}


As some long GRBs are associated with Type Ib,c supernovae, their progenitors are expected to be Wolf-Rayet stars.
Wolf-Rayet stars in our galaxy are observed to lose mass at a typical rate $\dot{M} \sim 10^{-5} M_{\odot} \ \mathrm{yr}^{-1}$ through strong winds of 
velocity $w \sim 10^8 \ \mathrm{cm \ s}^{-1}$ (e.g. \citealt{crowther_2007}). 
If the wind has constant $\dot{M}$, its density follows the $R^{-2}$ density profile,
\begin{equation}
\rho(R) = \frac{A}{R^2} \, ,
\end{equation}
where 
\begin{equation}
A= \frac{\dot{M}}{4\pi w} \simeq \left( \frac{\dot{M}}{10^{-5} M_{\odot} \ \mathrm{yr}^{-1}} \right)  \left. w_{8} \right.^{-1} \ A_0   
,  
 \quad 
A_0=5 \times10^{11} \ \mathrm{g \ cm^{-1}}  \, .
\end{equation}

As discussed in \sect{lf_dist}, the peak time $\tpz$ of the optical afterglow provides
an estimate of the ejecta Lorentz factor $\lfej\approx\lfbw$ if $\tpz \gg \tgamma$.
This estimate scales as $\rhodec^{-1/8}$.
On the other hand, an independent constraint on $\lfej$ can be derived by considering the Thomson opacity of the 
ejecta and its photosphere.\footnote{For GRBs with detected high-energy emission 
(above 100~MeV) another constraint on $\lfej$ could be derived from the requirement 
that the high-energy photons avoid $\gamma$-$\gamma$ absorption 
(e.g. \citealt{lithwick_2001, granot_2008, hascoet_2012}).
There are no such bursts in our sample.}
The photospheric radius $\rph$ (\eq{ph_radius}) implies
a minimum variability timescale $\dtobs\approx\rph/2 \lfej^2 c$. Thus, the observed $\dtobs$ is expected to satisfy the condition, 
\begin{equation}
\label{eqn_dtmin}
\frac{\dtobs}{1+z} \ga  \frac{\rph}{2 \lfej^2 c} 
\approx 5\,
\kappa_{0.2} \frac{1-\eta}{\eta} \lgammaNorm \lfej_{2}^{-5} \ \mathrm{ms} \, .
\end{equation} 
From this condition, using the observed $L_\gamma$ and $\dtobs$, 
one obtains a lower bound on the ejecta Lorentz factor.
Then, combining Equations (\ref{eq_lf_estimate}) and (\ref{eqn_dtmin}), 
an upper limit on $\rhodec$ can be derived. 
For a wind medium this limit translates into an upper bound on $A$,
\begin{equation}
\label{eq_amax}
A <  
A_{\max}\approx
10^{11} \left.\kappa_{0.2}\right.^{-4/5} \left(\frac{1-\eta}{\eta}\right)^{1/5} \left. \egammaNorm \right.^{1/5} 
\left( \frac{\dtobs}{1+z} \right)^{4/5} \left.\tgamma\right.^{4/5}
  \left. T_{\rm p,2} \right.^{-1} \ \mathrm{g \ cm^{-1}} \, .
\end{equation}
This upper limit is sensitive to the Thomson opacity $\kappa$, which may be 
significantly increased by $e^\pm$ creation in the ejecta.
The numerical value for $A_{\max}$ in Equation~(\ref{eq_amax}) is given for 
the lowest possible $\kappa=0.2$~cm$^2$~g$^{-1}$, which gives a conservative 
upper limit on $A$. \fig{fig_awind} shows the upper limits obtained  
for our GRB sample, where we used $\eta=0.5$, $\kappa_{0.2}=1$, and $\dtobs = 1$~s.
In some bursts, the inferred $A_{\max}$ is well below $A_0$.  
For GRB~060605, $A_{\max}$ is $\sim 6$ times smaller than $A_0$.
Even stronger limits could be derived with more sensitive detectors and more 
detailed analysis of variability timescales. GRBs that give the lowest $A_{\max}$
have relatively low luminosities, and poor photon statistics make it difficult
to see the true minimum $\dtobs$, which might be shorter than 1~s.  

\fig{fig_awind} suggests that at least some GRB progenitors are peculiar massive stars whose winds are
weaker than typical Wolf-Rayet stars observed in our galaxy.
This may be the result of a lower metallicity of the star (e.g. \citealt{vink_2001}), a property that seems to be preferred by GRB progenitors,
and agrees with observations of their host galaxies (e.g. \citealt{perley_2013}).
Weaker winds extract less angular momentum from the progenitor, leading to collapse with faster rotation, 
which is required for GRB central engines producing collimated jets (e.g. \citealt{woosley_2006}).
Note also that GRBs with bright optical afterglows (such as bursts in our sample) imply a selection against high-metallicity environment,
which tends to obscure the optical emission (e.g. \citealt{levesque_2013}).

The low wind density could also be explained by
a change in the stellar mass loss rate $\dot{M}$ shortly before the explosion.
The wind medium at a characteristic
radius $\rdec \sim 10^{16} \ \mathrm{cm}$ 
was ejected by the progenitor $\rdec/w \sim 3$ yr  
before the explosion.
The mass loss rates of Wolf-Rayet stars in the last few 
years of their lives are uncertain; $\dot{M}$ might decrease
as the star evolves toward the collapse.

\acknowledgements
This work was supported by NSF grant AST-1008334. 

\bibliographystyle{apj} 
\bibliography{main}

\begin{thebibliography}{94}
\expandafter\ifx\csname natexlab\endcsname\relax\def\natexlab#1{#1}\fi

\bibitem[{{Band}(2006)}]{band_2006}
{Band}, D.~L. 2006, \apj, 644, 378

\bibitem[{{Beloborodov}(2002)}]{beloborodov_2002}
{Beloborodov}, A.~M. 2002, \apj, 565, 808

\bibitem[{{Beloborodov}(2010)}]{beloborodov_2010}
---. 2010, \mnras, 407, 1033

\bibitem[{{Beloborodov}(2011)}]{beloborodov_2011}
---. 2011, \apj, 737, 68

\bibitem[{{Berger} {et~al.}(2008){Berger}, {Fox}, {Cucchiara}, \&
  {Cenko}}]{berger_2008}
{Berger}, E., {Fox}, D.~B., {Cucchiara}, A., \& {Cenko}, S.~B. 2008, GRB
  Coordinates Network, 8335, 1

\bibitem[{{Bloom} {et~al.}(2006){Bloom}, {Foley}, {Koceveki}, \&
  {Perley}}]{bloom_2006}
{Bloom}, J.~S., {Foley}, R.~J., {Koceveki}, D., \& {Perley}, D. 2006, GRB
  Coordinates Network, 5217, 1

\bibitem[{{Blustin} {et~al.}(2006){Blustin}, {Band}, {Barthelmy}, {Boyd},
  {Capalbi}, {Holland}, {Marshall}, {Mason}, {Perri}, {Poole}, {Roming},
  {Rosen}, {Schady}, {Still}, {Zhang}, {Angelini}, {Barbier}, {Beardmore},
  {Breeveld}, {Burrows}, {Cummings}, {Cannizzo}, {Campana}, {Chester},
  {Chincarini}, {Cominsky}, {Cucchiara}, {de Pasquale}, {Fenimore}, {Gehrels},
  {Giommi}, {Goad}, {Gronwall}, {Grupe}, {Hill}, {Hinshaw}, {Hunsberger},
  {Hurley}, {Ivanushkina}, {Kennea}, {Krimm}, {Kumar}, {Landsman}, {La Parola},
  {Markwardt}, {McGowan}, {M{\'e}sz{\'a}ros}, {Mineo}, {Moretti}, {Morgan},
  {Nousek}, {O'Brien}, {Osborne}, {Page}, {Page}, {Palmer}, {Parsons},
  {Rhoads}, {Romano}, {Sakamoto}, {Sato}, {Tagliaferri}, {Tueller}, {Wells}, \&
  {White}}]{blustin_2006}
{Blustin}, A.~J., {Band}, D., {Barthelmy}, S., {et~al.} 2006, \apj, 637, 901

\bibitem[{{Butler} {et~al.}(2006){Butler}, {Li}, {Perley}, {Huang}, {Urata},
  {Prochaska}, {Bloom}, {Filippenko}, {Foley}, {Kocevski}, {Chen}, {Qiu},
  {Kuo}, {Huang}, {Ip}, {Tamagawa}, {Onda}, {Tashiro}, {Makishima},
  {Nishihara}, \& {Sarugaku}}]{butler_2006}
{Butler}, N.~R., {Li}, W., {Perley}, D., {et~al.} 2006, \apj, 652, 1390

\bibitem[{{Cenko} {et~al.}(2013){Cenko}, {Kulkarni}, {Horesh}, {Corsi}, {Fox},
  {Carpenter}, {Frail}, {Nugent}, {Perley}, {Gruber}, {Gal-Yam}, {Groot},
  {Hallinan}, {Ofek}, {Rau}, {MacLeod}, {Miller}, {Bloom}, {Filippenko},
  {Kasliwal}, {Law}, {Morgan}, {Polishook}, {Poznanski}, {Quimby}, {Sesar},
  {Shen}, {Silverman}, \& {Sternberg}}]{cenko_2013}
{Cenko}, S.~B., {Kulkarni}, S.~R., {Horesh}, A., {et~al.} 2013, \apj, 769, 130

\bibitem[{{Chevalier} \& {Li}(2000)}]{chevalier_2000}
{Chevalier}, R.~A., \& {Li}, Z.-Y. 2000, \apj, 536, 195

\bibitem[{{Covino} {et~al.}(2010){Covino}, {Campana}, {Conciatore}, {D'Elia},
  {Palazzi}, {Th{\"o}ne}, {Vergani}, {Wiersema}, {Brusasca}, {Cucchiara},
  {Cobb}, {Fern{\'a}ndez-Soto}, {Kann}, {Malesani}, {Tanvir}, {Antonelli},
  {Bremer}, {Castro-Tirado}, {de Ugarte Postigo}, {Molinari}, {Nicastro},
  {Stefanon}, {Testa}, {Tosti}, {Vitali}, {Amati}, {Chapman}, {Conconi},
  {Cutispoto}, {Fynbo}, {Goldoni}, {Henriksen}, {Horne}, {Malaspina}, {Meurs},
  {Pian}, {Stella}, {Tagliaferri}, {Ward}, \& {Zerbi}}]{covino_2010}
{Covino}, S., {Campana}, S., {Conciatore}, M.~L., {et~al.} 2010, \aap, 521, A53

\bibitem[{{Crew} {et~al.}(2005){Crew}, {Ricker}, {Atteia}, {Kawai}, {Lamb},
  {Woosley}, {Arimoto}, {Donaghy}, {Fenimore}, {Galassi}, {Graziani}, {Kotoku},
  {Maetou}, {Matsuoka}, {Nakagawa}, {Sakamoto}, {Sato}, {Shirasaki}, {Suzuki},
  {Tamagawa}, {Tanaka}, {Yamamoto}, {Yoshida}, {Butler}, {Doty}, {Prigozhin},
  {Vanderspek}, {Villasenor}, {Jernigan}, {Levine}, {Azzibrouck}, {Braga},
  {Manchanda}, {Pizzichini}, {Boer}, {Olive}, {Dezalay}, \&
  {Hurley}}]{crew_2005}
{Crew}, G., {Ricker}, G., {Atteia}, J.-L., {et~al.} 2005, GRB Coordinates
  Network, 4021, 1

\bibitem[{{Crew} {et~al.}(2003){Crew}, {Lamb}, {Ricker}, {Atteia}, {Kawai},
  {Vanderspek}, {Villasenor}, {Doty}, {Prigozhin}, {Jernigan}, {Graziani},
  {Shirasaki}, {Sakamoto}, {Suzuki}, {Butler}, {Hurley}, {Tamagawa}, {Yoshida},
  {Matsuoka}, {Fenimore}, {Galassi}, {Barraud}, {Boer}, {Dezalay}, {Olive},
  {Levine}, {Monnelly}, {Martel}, {Morgan}, {Donaghy}, {Torii}, {Woosley},
  {Cline}, {Braga}, {Manchanda}, {Pizzichini}, {Takagishi}, \&
  {Yamauchi}}]{crew_2003}
{Crew}, G.~B., {Lamb}, D.~Q., {Ricker}, G.~R., {et~al.} 2003, \apj, 599, 387

\bibitem[{{Crowther}(2007)}]{crowther_2007}
{Crowther}, P.~A. 2007, \araa, 45, 177

\bibitem[{{Cucchiara} \& {Fox}(2008)}]{cucchiara_2008}
{Cucchiara}, A., \& {Fox}, D.~B. 2008, GRB Coordinates Network, 7654, 1

\bibitem[{{Cummings} {et~al.}(2006){Cummings}, {Barthelmy}, {Barbier},
  {Fenimore}, {Gehrels}, {Hullinger}, {Krimm}, {Koss}, {Markwardt}, {Palmer},
  {Parsons}, {Sakamoto}, {Sato}, {Stamatikos}, \& {Tueller}}]{cummings_2006}
{Cummings}, J., {Barthelmy}, S., {Barbier}, L., {et~al.} 2006, GRB Coordinates
  Network, 5124, 1

\bibitem[{{Cusumano} {et~al.}(2006){Cusumano}, {Mangano}, {Angelini},
  {Barthelmy}, {Beardmore}, {Burrows}, {Campana}, {Cannizzo}, {Capalbi},
  {Chincarini}, {Gehrels}, {Giommi}, {Goad}, {Hill}, {Kennea}, {Kobayashi}, {La
  Parola}, {Malesani}, {M{\'e}sz{\'a}ros}, {Mineo}, {Moretti}, {Nousek},
  {O'Brien}, {Osborne}, {Pagani}, {Page}, {Perri}, {Romano}, {Tagliaferri}, \&
  {Zhang}}]{cusumano_2006}
{Cusumano}, G., {Mangano}, V., {Angelini}, L., {et~al.} 2006, \apj, 639, 316

\bibitem[{{Daigne} \& {Mochkovitch}(1998)}]{daigne_1998}
{Daigne}, F., \& {Mochkovitch}, R. 1998, \mnras, 296, 275

\bibitem[{{Daigne} {et~al.}(2006){Daigne}, {Rossi}, \&
  {Mochkovitch}}]{daigne_2006}
{Daigne}, F., {Rossi}, E.~M., \& {Mochkovitch}, R. 2006, \mnras, 372, 1034

\bibitem[{{de Pasquale} \& {Cummings}(2006)}]{depasquale_2006}
{de Pasquale}, M., \& {Cummings}, J. 2006, GRB Coordinates Network, 5130, 1

\bibitem[{{Deng} {et~al.}(2009){Deng}, {Zheng}, {Zhai}, {Xin}, {Qiu},
  {Stefanescu}, {Pozanenko}, {Ibrahimov}, \& {Volnova}}]{deng_2009}
{Deng}, J., {Zheng}, W., {Zhai}, M., {et~al.} 2009, arXiv:0912.5435

\bibitem[{{Filgas} {et~al.}(2011){Filgas}, {Kr{\"u}hler}, {Greiner}, {Rau},
  {Palazzi}, {Klose}, {Schady}, {Rossi}, {Afonso}, {Antonelli}, {Clemens},
  {Covino}, {D'Avanzo}, {K{\"u}pc{\"u} Yolda{\c s}}, {Nardini}, {Nicuesa
  Guelbenzu}, {Olivares}, {Updike}, \& {Yolda{\c s}}}]{filgas_2011}
{Filgas}, R., {Kr{\"u}hler}, T., {Greiner}, J., {et~al.} 2011, \aap, 526, A113

\bibitem[{{Fugazza} {et~al.}(2005){Fugazza}, {Fiore}, {Patat}, {Ledoux},
  {D'Avanzo}, {Antonelli}, {Chincarini}, {Malesani}, {Covino}, {Tagliaferri},
  {Piranomonte}, \& {Stella}}]{fugazza_2005}
{Fugazza}, D., {Fiore}, F., {Patat}, N., {et~al.} 2005, GRB Coordinates
  Network, 3948, 1

\bibitem[{{Fynbo} {et~al.}(2005){Fynbo}, {Jensen}, {Hjorth}, {Wiersema},
  {Starling}, {Vreeswijk}, {Rol}, {Levan}, {Ellison}, \&
  {Masetti}}]{fynbo_2005}
{Fynbo}, J.~P.~U., {Jensen}, B.~L., {Hjorth}, J., {et~al.} 2005, GRB
  Coordinates Network, 3176, 1

\bibitem[{{Gehrels} {et~al.}(2008){Gehrels}, {Barthelmy}, {Burrows},
  {Cannizzo}, {Chincarini}, {Fenimore}, {Kouveliotou}, {O'Brien}, {Palmer},
  {Racusin}, {Roming}, {Sakamoto}, {Tueller}, {Wijers}, \&
  {Zhang}}]{gehrels_2008}
{Gehrels}, N., {Barthelmy}, S.~D., {Burrows}, D.~N., {et~al.} 2008, \apj, 689,
  1161

\bibitem[{{Gendre} {et~al.}(2010){Gendre}, {Klotz}, {Palazzi}, {Kr{\"u}hler},
  {Covino}, {Afonso}, {Antonelli}, {Atteia}, {D'Avanzo}, {Bo{\"e}r}, {Greiner},
  \& {Klose}}]{gendre_2010}
{Gendre}, B., {Klotz}, A., {Palazzi}, E., {et~al.} 2010, \mnras, 405, 2372

\bibitem[{{Genet} {et~al.}(2007){Genet}, {Daigne}, \&
  {Mochkovitch}}]{genet_2007}
{Genet}, F., {Daigne}, F., \& {Mochkovitch}, R. 2007, \mnras, 381, 732

\bibitem[{{Ghirlanda} {et~al.}(2003){Ghirlanda}, {Celotti}, \&
  {Ghisellini}}]{ghirlanda_2003}
{Ghirlanda}, G., {Celotti}, A., \& {Ghisellini}, G. 2003, \aap, 406, 879

\bibitem[{{Ghirlanda} {et~al.}(2012){Ghirlanda}, {Nava}, {Ghisellini},
  {Celotti}, {Burlon}, {Covino}, \& {Melandri}}]{ghirlanda_2012}
{Ghirlanda}, G., {Nava}, L., {Ghisellini}, G., {et~al.} 2012, \mnras, 420, 483

\bibitem[{{Goldstein} {et~al.}(2012){Goldstein}, {Burgess}, {Preece}, {Briggs},
  {Guiriec}, {van der Horst}, {Connaughton}, {Wilson-Hodge}, {Paciesas},
  {Meegan}, {von Kienlin}, {Bhat}, {Bissaldi}, {Chaplin}, {Diehl}, {Fishman},
  {Fitzpatrick}, {Foley}, {Gibby}, {Giles}, {Greiner}, {Gruber}, {Kippen},
  {Kouveliotou}, {McBreen}, {McGlynn}, {Rau}, \& {Tierney}}]{goldstein_2012}
{Goldstein}, A., {Burgess}, J.~M., {Preece}, R.~D., {et~al.} 2012, \apjs, 199,
  19

\bibitem[{{Golenetskii} {et~al.}(2005){Golenetskii}, {Aptekar}, {Mazets},
  {Pal'Shin}, {Frederiks}, \& {Cline}}]{golenetskii_2005}
{Golenetskii}, S., {Aptekar}, R., {Mazets}, E., {et~al.} 2005, GRB Coordinates
  Network, 4238, 1

\bibitem[{{Goodman}(1986)}]{goodman_1986}
{Goodman}, J. 1986, \apjl, 308, L47

\bibitem[{{Granot} {et~al.}(2008){Granot}, {Cohen-Tanugi}, \& {do Couto e
  Silva}}]{granot_2008}
{Granot}, J., {Cohen-Tanugi}, J., \& {do Couto e Silva}, E. 2008, \apj, 677, 92

\bibitem[{{Grupe} {et~al.}(2009){Grupe}, {Marshall}, {Cummings}, {Siegel},
  {Vetere}, {Barthelmy}, {Burrows}, {Roming}, \& {Gehrels}}]{grupe_2009}
{Grupe}, D., {Marshall}, F.~E., {Cummings}, J.~R., {et~al.} 2009, GCN Report,
  260, 1

\bibitem[{{Guidorzi} {et~al.}(2008){Guidorzi}, {Stamatikos}, {Landsman},
  {Barthelmy}, {Burrows}, {Roming}, \& {Gehrels}}]{guidorzi_2008}
{Guidorzi}, C., {Stamatikos}, M., {Landsman}, W., {et~al.} 2008, GCN Report,
  139, 1

\bibitem[{{Guidorzi} {et~al.}(2011){Guidorzi}, {Kobayashi}, {Perley},
  {Vianello}, {Bloom}, {Chandra}, {Kann}, {Li}, {Mundell}, {Pozanenko},
  {Prochaska}, {Antoniuk}, {Bersier}, {Filippenko}, {Frail}, {Gomboc},
  {Klunko}, {Melandri}, {Mereghetti}, {Morgan}, {O'Brien}, {Rumyantsev},
  {Smith}, {Steele}, {Tanvir}, \& {Volnova}}]{guidorzi_2011}
{Guidorzi}, C., {Kobayashi}, S., {Perley}, D.~A., {et~al.} 2011, \mnras, 417,
  2124

\bibitem[{{Hasco{\"e}t} {et~al.}(2012){Hasco{\"e}t}, {Daigne}, {Mochkovitch},
  \& {Vennin}}]{hascoet_2012}
{Hasco{\"e}t}, R., {Daigne}, F., {Mochkovitch}, R., \& {Vennin}, V. 2012,
  \mnras, 421, 525

\bibitem[{{Huang} {et~al.}(2012){Huang}, {Urata}, {Tung}, {Lin}, {Xin},
  {Yoshida}, {Zheng}, {Akerlof}, {Wang}, {Ip}, {Lehner}, {Bianco}, {Kawai},
  {Kuroda}, {Marshall}, {Schwamb}, {Qiu}, {Wang}, {Wen}, {Wei}, {Yanagisawa},
  \& {Zhang}}]{huang_2012}
{Huang}, K.~Y., {Urata}, Y., {Tung}, Y.~H., {et~al.} 2012, \apj, 748, 44

\bibitem[{{Huang} {et~al.}(2002){Huang}, {Dai}, \& {Lu}}]{huang_2002}
{Huang}, Y.~F., {Dai}, Z.~G., \& {Lu}, T. 2002, \mnras, 332, 735

\bibitem[{{Hunsberger} {et~al.}(2005){Hunsberger}, {Marshall}, {Holland},
  {et~al.}}]{hunsberger_2005}
{Hunsberger}, S.~D., {Marshall}, F., {Holland}, S.~T., {et~al.} 2005, GCN
  Report, 4041, 1

\bibitem[{{Jelinek} {et~al.}(2008){Jelinek}, {Kubanek}, {Gorosabel},
  {Castro-Tirado}, {Aceituno}, {Sabau-Graziati}, {de Ugarte Postigo}, {Hudec},
  {Perez-Gonzalez}, \& {Zamorano}}]{jelinek_2008}
{Jelinek}, M., {Kubanek}, P., {Gorosabel}, J., {et~al.} 2008, GRB Coordinates
  Network, 7648, 1

\bibitem[{{Kaneko} {et~al.}(2006){Kaneko}, {Preece}, {Briggs}, {Paciesas},
  {Meegan}, \& {Band}}]{kaneko_2006}
{Kaneko}, Y., {Preece}, R.~D., {Briggs}, M.~S., {et~al.} 2006, \apjs, 166, 298

\bibitem[{{Krimm} {et~al.}(2005{\natexlab{a}}){Krimm}, {Barbier}, {Barthelmy},
  {Cannizzo}, {Cominsky}, {Cummings}, {Fenimore}, {Gehrels}, {Hullinger},
  {Markwardt}, {Palmer}, {Parsons}, {Sakamoto}, {Sato}, \&
  {Tueller}}]{krimm_2005}
{Krimm}, H., {Barbier}, L., {Barthelmy}, S., {et~al.} 2005{\natexlab{a}}, GRB
  Coordinates Network, 4020, 1

\bibitem[{{Krimm} {et~al.}(2005{\natexlab{b}}){Krimm}, {Ajello}, {Barbier},
  {Barthelmy}, {Cummings}, {Fenimore}, {Fink}, {Gehrels}, {Hullinger},
  {Markwardt}, {Palmer}, {Parsons}, {Sakamoto}, {Sato}, \&
  {Tueller}}]{krimm_2005b}
{Krimm}, H., {Ajello}, M., {Barbier}, L., {et~al.} 2005{\natexlab{b}}, GRB
  Coordinates Network, 4260, 1

\bibitem[{{Landsman} \& {Guidorzi}(2008)}]{landsman_2008}
{Landsman}, W.~B., \& {Guidorzi}, C. 2008, GRB Coordinates Network, 7660, 1

\bibitem[{{Levesque}(2013)}]{levesque_2013}
{Levesque}, E.~M. 2013, arXiv:1302.4741

\bibitem[{{Li} {et~al.}(2003){Li}, {Filippenko}, {Chornock}, \&
  {Jha}}]{li_2003}
{Li}, W., {Filippenko}, A.~V., {Chornock}, R., \& {Jha}, S. 2003, \apjl, 586,
  L9

\bibitem[{{Liang} {et~al.}(2010){Liang}, {Yi}, {Zhang}, {L{\"u}}, {Zhang}, \&
  {Zhang}}]{liang_2010}
{Liang}, E.-W., {Yi}, S.-X., {Zhang}, J., {et~al.} 2010, \apj, 725, 2209

\bibitem[{{Liang} {et~al.}(2013){Liang}, {Li}, {Gao}, {Zhang}, {Liang}, {Wu},
  {Yi}, {Dai}, {Tang}, {Chen}, {L{\"u}}, {Zhang}, {Lu}, {L{\"u}}, \&
  {Wei}}]{liang_2012}
{Liang}, E.-W., {Li}, L., {Gao}, H., {et~al.} 2013, \apj, 774, 13

\bibitem[{{Lithwick} \& {Sari}(2001)}]{lithwick_2001}
{Lithwick}, Y., \& {Sari}, R. 2001, \apj, 555, 540

\bibitem[{{L{\"u}} {et~al.}(2012){L{\"u}}, {Zou}, {Lei}, {Zhang}, {Wu}, {Wang},
  {Liang}, \& {L{\"u}}}]{lue_2012}
{L{\"u}}, J., {Zou}, Y.-C., {Lei}, W.-H., {et~al.} 2012, \apj, 751, 49

\bibitem[{{Markwardt} {et~al.}(2008){Markwardt}, {Barthelmy}, {Baumgartner},
  {Cummings}, {Fenimore}, {Gehrels}, {Krimm}, {McLean}, {Palmer}, {Parsons},
  {Sakamoto}, {Sato}, {Stamatikos}, {Tueller}, \& {Ukwatta}}]{markwardt_2008}
{Markwardt}, C.~M., {Barthelmy}, S.~D., {Baumgartner}, W.~H., {et~al.} 2008,
  GRB Coordinates Network, 8338, 1

\bibitem[{{Martin-Carrillo} {et~al.}(2008){Martin-Carrillo}, {Hanlon},
  {McGlynn}, {Foley}, {McBreen}, {Melady}, {French}, {Kubanek}, {Ferrero},
  {McBreen}, {Molkov}, {Preece}, \& {von Kienlin}}]{martin_2008}
{Martin-Carrillo}, A., {Hanlon}, L., {McGlynn}, S., {et~al.} 2008, in Proc. 7th
  INTEGRAL Workshop, 2008 September 8--11 (Proceedings of Science: Copenhagen,
  Denmark), 16

\bibitem[{{Meszaros} \& {Rees}(1997)}]{meszaros_1997}
{Meszaros}, P., \& {Rees}, M.~J. 1997, \apj, 476, 232

\bibitem[{{Molinari} {et~al.}(2007){Molinari}, {Vergani}, {Malesani}, {Covino},
  {D'Avanzo}, {Chincarini}, {Zerbi}, {Antonelli}, {Conconi}, {Testa}, {Tosti},
  {Vitali}, {D'Alessio}, {Malaspina}, {Nicastro}, {Palazzi}, {Guetta},
  {Campana}, {Goldoni}, {Masetti}, {Meurs}, {Monfardini}, {Norci}, {Pian},
  {Piranomonte}, {Rizzuto}, {Stefanon}, {Stella}, {Tagliaferri}, {Ward},
  {Ihle}, {Gonzalez}, {Pizarro}, {Sinclaire}, \& {Valenzuela}}]{molinari_2007}
{Molinari}, E., {Vergani}, S.~D., {Malesani}, D., {et~al.} 2007, \aap, 469, L13

\bibitem[{{Nava} {et~al.}(2012){Nava}, {Salvaterra}, {Ghirlanda}, {Ghisellini},
  {Campana}, {Covino}, {Cusumano}, {D'Avanzo}, {D'Elia}, {Fugazza}, {Melandri},
  {Sbarufatti}, {Vergani}, \& {Tagliaferri}}]{nava_2012}
{Nava}, L., {Salvaterra}, R., {Ghirlanda}, G., {et~al.} 2012, \mnras, 421, 1256

\bibitem[{{Oksanen} \& {Hentunen}(2008)}]{oksanen_2008}
{Oksanen}, A., \& {Hentunen}, V.-P. 2008, GRB Coordinates Network, 7657, 1

\bibitem[{{Paczynski}(1986)}]{paczynski_1986}
{Paczynski}, B. 1986, \apjl, 308, L43

\bibitem[{{Page} {et~al.}(2011){Page}, {Starling}, {Fitzpatrick}, {Pandey},
  {Osborne}, {Schady}, {McBreen}, {Campana}, {Ukwatta}, {Pagani}, {Beardmore},
  \& {Evans}}]{page_2011}
{Page}, K.~L., {Starling}, R.~L.~C., {Fitzpatrick}, G., {et~al.} 2011, \mnras,
  416, 2078

\bibitem[{{Pandey} {et~al.}(2003){Pandey}, {Anupama}, {Sagar}, {Bhattacharya},
  {Castro-Tirado}, {Sahu}, {Parihar}, \& {Prabhu}}]{pandey_2003}
{Pandey}, S.~B., {Anupama}, G.~C., {Sagar}, R., {et~al.} 2003, \aap, 408, L21

\bibitem[{{Perley} {et~al.}(2008){Perley}, {Li}, {Chornock}, {Prochaska},
  {Butler}, {Chandra}, {Pollack}, {Bloom}, {Filippenko}, {Swan}, {Yuan},
  {Akerlof}, {Auger}, {Cenko}, {Chen}, {Fassnacht}, {Fox}, {Frail},
  {Johansson}, {McKay}, {Le Mignant}, {Modjaz}, {Rujopakarn}, {Russel},
  {Skinner}, {Smith}, {Smith}, {van Dam}, \& {Yost}}]{perley_2008}
{Perley}, D.~A., {Li}, W., {Chornock}, R., {et~al.} 2008, \apj, 688, 470

\bibitem[{{Perley} {et~al.}(2013){Perley}, {Levan}, {Tanvir}, {Cenko}, {Bloom},
  {Hjorth}, {Kruehler}, {Filippenko}, {Fruchter}, {Fynbo}, {Jakobsson},
  {Kalirai}, {Milvang-Jensen}, {Morgan}, {Prochaska}, \&
  {Silverman}}]{perley_2013}
{Perley}, D.~A., {Levan}, A.~J., {Tanvir}, N.~R., {et~al.} 2013,
  arXiv:1301.5903

\bibitem[{{Piran}(2004)}]{piran_2004}
{Piran}, T. 2004, Reviews of Modern Physics, 76, 1143

\bibitem[{{Porciani} \& {Madau}(2001)}]{porciani_2001}
{Porciani}, C., \& {Madau}, P. 2001, \apj, 548, 522

\bibitem[{{Quimby} {et~al.}(2006){Quimby}, {Rykoff}, {Yost}, {Aharonian},
  {Akerlof}, {Alatalo}, {Ashley}, {G{\"o}{\u g}{\"u}{\c s}}, {G{\"u}ver},
  {Horns}, {Kehoe}, {K{$\iota$}z{$\iota$}lo{\u g}lu}, {Mckay}, {{\"O}zel},
  {Phillips}, {Schaefer}, {Smith}, {Swan}, {Vestrand}, {Wheeler}, \&
  {Wren}}]{quimby_2006}
{Quimby}, R.~M., {Rykoff}, E.~S., {Yost}, S.~A., {et~al.} 2006, \apj, 640, 402

\bibitem[{{Racusin} {et~al.}(2008){Racusin}, {Karpov}, {Sokolowski}, {Granot},
  {Wu}, {Pal'Shin}, {Covino}, {van der Horst}, {Oates}, {Schady}, {Smith},
  {Cummings}, {Starling}, {Piotrowski}, {Zhang}, {Evans}, {Holland}, {Malek},
  {Page}, {Vetere}, {Margutti}, {Guidorzi}, {Kamble}, {Curran}, {Beardmore},
  {Kouveliotou}, {Mankiewicz}, {Melandri}, {O'Brien}, {Page}, {Piran},
  {Tanvir}, {Wrochna}, {Aptekar}, {Barthelmy}, {Bartolini}, {Beskin}, {Bondar},
  {Bremer}, {Campana}, {Castro-Tirado}, {Cucchiara}, {Cwiok}, {D'Avanzo},
  {D'Elia}, {Della Valle}, {de Ugarte Postigo}, {Dominik}, {Falcone}, {Fiore},
  {Fox}, {Frederiks}, {Fruchter}, {Fugazza}, {Garrett}, {Gehrels},
  {Golenetskii}, {Gomboc}, {Gorosabel}, {Greco}, {Guarnieri}, {Immler},
  {Jelinek}, {Kasprowicz}, {La Parola}, {Levan}, {Mangano}, {Mazets},
  {Molinari}, {Moretti}, {Nawrocki}, {Oleynik}, {Osborne}, {Pagani}, {Pandey},
  {Paragi}, {Perri}, {Piccioni}, {Ramirez-Ruiz}, {Roming}, {Steele}, {Strom},
  {Testa}, {Tosti}, {Ulanov}, {Wiersema}, {Wijers}, {Winters}, {Zarnecki},
  {Zerbi}, {M{\'e}sz{\'a}ros}, {Chincarini}, \& {Burrows}}]{racusin_2008}
{Racusin}, J.~L., {Karpov}, S.~V., {Sokolowski}, M., {et~al.} 2008, \nat, 455,
  183

\bibitem[{{Rees} \& {Meszaros}(1994)}]{rees_1994}
{Rees}, M.~J., \& {Meszaros}, P. 1994, \apjl, 430, L93

\bibitem[{{Ryde}(2004)}]{ryde_2004}
{Ryde}, F. 2004, \apj, 614, 827

\bibitem[{{Rykoff} {et~al.}(2005){Rykoff}, {Yost}, {Krimm}, {Aharonian},
  {Akerlof}, {Alatalo}, {Ashley}, {Barthelmy}, {Gehrels}, {G{\"o}{\v g}{\"u}{\c
  s}}, {G{\"u}ver}, {Horns}, {K{\i}z{\i}lo{\v g}lu}, {McKay}, {{\"O}zel},
  {Phillips}, {Quimby}, {Rujopakarn}, {Schaefer}, {Smith}, {Swan}, {Vestrand},
  {Wheeler}, \& {Wren}}]{rykoff_2005b}
{Rykoff}, E.~S., {Yost}, S.~A., {Krimm}, H.~A., {et~al.} 2005, \apjl, 631, L121

\bibitem[{{Rykoff} {et~al.}(2009){Rykoff}, {Aharonian}, {Akerlof}, {Ashley},
  {Barthelmy}, {Flewelling}, {Gehrels}, {G{\"o}{\v g}{\"u}{\c s}}, {G{\"u}ver},
  {Kizilo{\v g}lu}, {Krimm}, {McKay}, {{\"O}zel}, {Phillips}, {Quimby},
  {Rowell}, {Rujopakarn}, {Schaefer}, {Smith}, {Vestrand}, {Wheeler}, {Wren},
  {Yuan}, \& {Yost}}]{rykoff_2009}
{Rykoff}, E.~S., {Aharonian}, F., {Akerlof}, C.~W., {et~al.} 2009, \apj, 702,
  489

\bibitem[{{Sakamoto} {et~al.}(2005){Sakamoto}, {Barthelmy}, {Barbier},
  {Cummings}, {Fenimore}, {Gehrels}, {Hullinger}, {Krimm}, {Markwardt},
  {Palmer}, {Parsons}, {Sato}, {Suzuki}, {Tashiro}, \&
  {Tueller}}]{sakamoto_2005b}
{Sakamoto}, T., {Barthelmy}, S., {Barbier}, L., {et~al.} 2005, GRB Coordinates
  Network, 3173, 1

\bibitem[{{Salvaterra} {et~al.}(2012){Salvaterra}, {Campana}, {Vergani},
  {Covino}, {D'Avanzo}, {Fugazza}, {Ghirlanda}, {Ghisellini}, {Melandri},
  {Nava}, {Sbarufatti}, {Flores}, {Piranomonte}, \&
  {Tagliaferri}}]{salvaterra_2012}
{Salvaterra}, R., {Campana}, S., {Vergani}, S.~D., {et~al.} 2012, \apj, 749, 68

\bibitem[{{Sari} \& {Piran}(1999)}]{sari_1999b}
{Sari}, R., \& {Piran}, T. 1999, \aaps, 138, 537

\bibitem[{{Sari} {et~al.}(1998){Sari}, {Piran}, \& {Narayan}}]{sari_1998}
{Sari}, R., {Piran}, T., \& {Narayan}, R. 1998, \apjl, 497, L17

\bibitem[{{Sato} {et~al.}(2005){Sato}, {Barbier}, {Barthelmy}, {Boyd},
  {Cummings}, {Hullinger}, {Fenimore}, {Gehrels}, {Krimm}, {Markwardt},
  {Palmer}, {Parsons}, {Sakamoto}, {Tueller}, \& {Voges}}]{sato_2005}
{Sato}, G., {Barbier}, L., {Barthelmy}, S., {et~al.} 2005, GRB Coordinates
  Network, 3951, 1

\bibitem[{{Sollerman} {et~al.}(2007){Sollerman}, {Fynbo}, {Gorosabel},
  {Halpern}, {Hjorth}, {Jakobsson}, {Mirabal}, {Watson}, {Xu}, {Castro-Tirado},
  {F{\'e}ron}, {Jaunsen}, {Jel{\'{\i}}nek}, {Jensen}, {Kann}, {Ovaldsen},
  {Pozanenko}, {Stritzinger}, {Th{\"o}ne}, {de Ugarte Postigo}, {Guziy},
  {Ibrahimov}, {J{\"a}rvinen}, {Levan}, {Rumyantsev}, \&
  {Tanvir}}]{sollerman_2007}
{Sollerman}, J., {Fynbo}, J.~P.~U., {Gorosabel}, J., {et~al.} 2007, \aap, 466,
  839

\bibitem[{{Spruit} {et~al.}(2001){Spruit}, {Daigne}, \&
  {Drenkhahn}}]{spruit_2001}
{Spruit}, H.~C., {Daigne}, F., \& {Drenkhahn}, G. 2001, \aap, 369, 694

\bibitem[{{Stamatikos} {et~al.}(2007){Stamatikos}, {Barthelmy}, {Cummings},
  {Fenimore}, {Gehrels}, {Krimm}, {Markwardt}, {Palmer}, {Sakamoto}, {Sato},
  {Stroh}, {Tueller}, \& {Ukwatta}}]{stamatikos_2007}
{Stamatikos}, M., {Barthelmy}, S.~D., {Cummings}, J., {et~al.} 2007, GRB
  Coordinates Network, 7029, 1

\bibitem[{{Stamatikos} {et~al.}(2009){Stamatikos}, {Cummings}, {Evans},
  {Gronwall}, {Guidorzi}, {Markwardt}, {O'Brien}, {Page}, {Palmer}, \&
  {Starling}}]{stamatikos_2009}
{Stamatikos}, M., {Cummings}, J.~R., {Evans}, P.~A., {et~al.} 2009, GRB
  Coordinates Network, 9768, 1

\bibitem[{{Starling} {et~al.}(2009){Starling}, {Rol}, {van der Horst}, {Yoon},
  {Pal'Shin}, {Ledoux}, {Page}, {Fynbo}, {Wiersema}, {Tanvir}, {Jakobsson},
  {Guidorzi}, {Curran}, {Levan}, {O'Brien}, {Osborne}, {Svinkin}, {de Ugarte
  Postigo}, {Oosting}, \& {Howarth}}]{starling_2009}
{Starling}, R.~L.~C., {Rol}, E., {van der Horst}, A.~J., {et~al.} 2009, \mnras,
  400, 90

\bibitem[{{Thompson} \& {Madau}(2000)}]{thompson_2000}
{Thompson}, C., \& {Madau}, P. 2000, \apj, 538, 105

\bibitem[{{Uehara} {et~al.}(2012){Uehara}, {Toma}, {Kawabata}, {Chiyonobu},
  {Fukazawa}, {Ikejiri}, {Inoue}, {Itoh}, {Komatsu}, {Miyamoto}, {Mizuno},
  {Nagae}, {Nakaya}, {Ohsugi}, {Sakimoto}, {Sasada}, {Tanaka}, {Uemura},
  {Yamanaka}, {Yamashita}, {Yamazaki}, \& {Yoshida}}]{uehara_2012}
{Uehara}, T., {Toma}, K., {Kawabata}, K.~S., {et~al.} 2012, \apjl, 752, L6

\bibitem[{{Uhm} \& {Beloborodov}(2007)}]{uhm_2007}
{Uhm}, Z.~L., \& {Beloborodov}, A.~M. 2007, \apjl, 665, L93

\bibitem[{{Ukwatta} {et~al.}(2008){Ukwatta}, {Barthelmy}, {Baumgartner},
  {Cummings}, {Fenimore}, {Gehrels}, {Krimm}, {Markwardt}, {Palmer}, {Parsons},
  {Sakamoto}, {Sato}, {Stamatikos}, \& {Tueller}}]{ukwatta_2008}
{Ukwatta}, T.~N., {Barthelmy}, S.~D., {Baumgartner}, W.~H., {et~al.} 2008, GRB
  Coordinates Network, 8599, 1

\bibitem[{{Vink} {et~al.}(2001){Vink}, {de Koter}, \& {Lamers}}]{vink_2001}
{Vink}, J.~S., {de Koter}, A., \& {Lamers}, H.~J.~G.~L.~M. 2001, \aap, 369, 574

\bibitem[{{Wanderman} \& {Piran}(2010)}]{wanderman_2010}
{Wanderman}, D., \& {Piran}, T. 2010, \mnras, 406, 1944

\bibitem[{{Wei} \& {Gao}(2003)}]{wei_2003}
{Wei}, D.~M., \& {Gao}, W.~H. 2003, \mnras, 345, 743

\bibitem[{{Wiersema} {et~al.}(2008){Wiersema}, {van der Horst}, {Kann}, {Rol},
  {Starling}, {Curran}, {Gorosabel}, {Levan}, {Fynbo}, {de Ugarte Postigo},
  {Wijers}, {Castro-Tirado}, {Guziy}, {Hornstrup}, {Hjorth}, {Jel{\'{\i}}nek},
  {Jensen}, {Kidger}, {Mart{\'{\i}}n-Luis}, {Tanvir}, {Tristram}, \&
  {Vreeswijk}}]{wiersema_2008}
{Wiersema}, K., {van der Horst}, A.~J., {Kann}, D.~A., {et~al.} 2008, \aap,
  481, 319

\bibitem[{{Wiersema} {et~al.}(2012){Wiersema}, {Curran}, {Kr{\"u}hler},
  {Melandri}, {Rol}, {Starling}, {Tanvir}, {van der Horst}, {Covino}, {Fynbo},
  {Goldoni}, {Gorosabel}, {Hjorth}, {Klose}, {Mundell}, {O'Brien}, {Palazzi},
  {Wijers}, {D'Elia}, {Evans}, {Filgas}, {Gomboc}, {Greiner}, {Guidorzi},
  {Kaper}, {Kobayashi}, {Kouveliotou}, {Levan}, {Rossi}, {Rowlinson}, {Steele},
  {de Ugarte Postigo}, \& {Vergani}}]{wiersema_2012}
{Wiersema}, K., {Curran}, P.~A., {Kr{\"u}hler}, T., {et~al.} 2012, \mnras, 426,
  2

\bibitem[{{Woosley} \& {Heger}(2006)}]{woosley_2006}
{Woosley}, S.~E., \& {Heger}, A. 2006, \apj, 637, 914

\bibitem[{{Wren} {et~al.}(2008){Wren}, {Vestrand}, {Wozniak}, {Davis}, \&
  {Norman}}]{wren_2008}
{Wren}, J., {Vestrand}, W.~T., {Wozniak}, P.~R., {Davis}, H., \& {Norman}, B.
  2008, GRB Coordinates Network, 8337, 1

\bibitem[{{Yonetoku} {et~al.}(2004){Yonetoku}, {Murakami}, {Nakamura},
  {Yamazaki}, {Inoue}, \& {Ioka}}]{yonetoku_2004}
{Yonetoku}, D., {Murakami}, T., {Nakamura}, T., {et~al.} 2004, \apj, 609, 935

\bibitem[{{Yuan} {et~al.}(2010){Yuan}, {Schady}, {Racusin}, {Willingale},
  {Kr{\"u}hler}, {O'Brien}, {Greiner}, {Oates}, {Rykoff}, {Aharonian},
  {Akerlof}, {Ashley}, {Barthelmy}, {Filgas}, {Flewelling}, {Gehrels},
  {G{\"o}{\u g}{\"u}{\c s}}, {G{\"u}ver}, {Horns}, {K{\i}z{\i}lo{\v g}lu},
  {Krimm}, {McKay}, {{\"O}zel}, {Phillips}, {Quimby}, {Rowell}, {Rujopakarn},
  {Schaefer}, {Vestrand}, {Wheeler}, \& {Wren}}]{yuan_2010}
{Yuan}, F., {Schady}, P., {Racusin}, J.~L., {et~al.} 2010, \apj, 711, 870

\bibitem[{{Zaninoni} {et~al.}(2013){Zaninoni}, {Bernardini}, {Margutti},
  {Oates}, \& {Chincarini}}]{zaninoni_2013}
{Zaninoni}, E., {Bernardini}, M.~G., {Margutti}, R., {Oates}, S., \&
  {Chincarini}, G. 2013, \aap, 557, A12

\end{thebibliography}

\end{document}